\documentclass[structabstract]{aa}  

\usepackage{graphicx}
\usepackage{txfonts}

\usepackage{units}
\usepackage{varioref}
\usepackage[breaklinks]{hyperref}

\usepackage{natbib}
\bibpunct{(}{)}{;}{a}{}{,} % to follow the A&A style

\newcommand{\Rn}{\mathbb{R}^n}

\newcommand{\dd}{\ensuremath{\mathrm d}}
\newcommand{\e}{\ensuremath{\mathrm e}}
\newcommand{\im}{\ensuremath{\mathrm i}}

\newcommand{\atanh}{\ensuremath{\mathrm{atanh}}}

\newcommand{\transpose}[1]{\ensuremath{#1^{\mathrm T}}}

\newcommand{\ie}{i.e.\ }
\newcommand{\eg}{e.g.\ }

\newcommand{\riu}{\ensuremath{r_{i\mathrm{u}}}}
\newcommand{\ril}{\ensuremath{r_{i\mathrm{l}}}}
\newcommand{\rnu}{\ensuremath{r_{n\mathrm{u}}}}
\newcommand{\rnl}{\ensuremath{r_{n\mathrm{l}}}}
\newcommand{\Deltai}{\ensuremath{(\riu-\ril)}}
\newcommand{\DKL}{\ensuremath{D_\mathrm{KL}}}

\newcommand{\Deltaint}{\ensuremath{\Delta_\mathrm{int}}}

\newcommand{\Cnzero}{\ensuremath{C_{n0}}}
\newcommand{\Cnone}{\ensuremath{C_{n1}}}
\newcommand{\Cntwo}{\ensuremath{C_{n2}}}
\newcommand{\Cmzero}{\ensuremath{C_{m0}}}

\newcommand{\Cmtwo}{\ensuremath{C_{m2}}}
\newcommand{\Ckzero}{\ensuremath{C_{k0}}}

\newcommand{\Cktwo}{\ensuremath{C_{k2}}}

\newcommand{\cov}{\ensuremath{\mathrm{cov}}}
\newcommand{\SH}{SH2009}
\newcommand{\KS}{KS2011}

\labelformat{equation}{\textup{(#1)}}

\begin{document}

   \title{A quasi-Gaussian approximation for the probability distribution of correlation functions}

   \author{P. Wilking \and P. Schneider}

   \institute{Argelander-Institut f\"ur Astronomie, Universit\"at Bonn, Auf dem H\"ugel 71, 53121 Bonn, Germany\\ \email{[pwilking;peter]@astro.uni-bonn.de}}

   \date{Received 17/04/2013; accepted 01/07/2013}

   \abstract
   % context heading (optional)
   {Whenever correlation functions are used for inference about cosmological parameters in the context of a Bayesian analysis, the likelihood function of correlation functions needs to be known. Usually, it is approximated as a multivariate Gaussian, though this is not necessarily a good approximation.}
   % aims heading (mandatory)
   {We show how to calculate a better approximation for the probability distribution of correlation functions of one-dimensional random fields, which we call ``quasi-Gaussian''.}
   % methods heading (mandatory)
   {Using the exact univariate PDF as well as constraints on correlation functions previously derived, we transform the correlation functions to an unconstrained variable for which the Gaussian approximation is well justified. From this Gaussian in the transformed space, we obtain the quasi-Gaussian PDF. The two approximations for the probability distributions are compared to the ``true'' distribution as obtained from simulations. Additionally, we test how the new approximation performs when used as likelihood in a toy-model Bayesian analysis.}
   % results heading (mandatory)
   {The quasi-Gaussian PDF agrees very well with the PDF obtained from simulations; in particular, it provides a significantly better description than a straightforward copula approach. In a simple toy-model likelihood analysis, it yields noticeably different results than the Gaussian likelihood, indicating its possible impact on cosmological parameter estimation.}
  % conclusions heading (optional)
   {}


\maketitle

%________________________________________________________________

\section{Introduction}
\label{sec:intro}
Correlation functions are an important tool in cosmology, since they constitute one standard way of using observations to constrain cosmological parameters. The two-point correlation function $\xi(\vec{x})$ obtained from observational data is usually analyzed in the framework of Bayesian statistics: Using Bayes' Theorem
\begin{equation}
 p(\vec{\theta}|\xi)=\frac{\mathcal{L}(\vec{\theta})\cdot p(\vec{\theta})}{p(\xi)},
 \label{eq:bayes}
\end{equation}
the measured correlation function $\xi$ is used to quantify how the optimal parameters $\vec{\theta}$ of the cosmological model change. It is inherent to Bayesian methods that we need to ``plug in'' the probability distribution function (PDF) of the data involved. In the case of correlation functions, this likelihood $\mathcal{L}(\vec{\theta})\equiv p(\xi|\vec{\theta})$ is usually assumed to be a Gaussian. For example in \cite{bib_COBE_likelihood_corr_function}, the authors perform an analysis of the angular correlation function of the cosmic microwave background (as calculated from COBE data), stating that a Gaussian is a good approximation for its likelihood, at least near the peak. Also in common methods of baryon acoustic oscillations (BAO) detection, a Gaussian distribution of correlation functions is usually assumed (see \eg \citealt{bib_bao_labatie}).
While this may be a good approximation in some cases, there are strong hints that there may in other cases be strong deviations from a Gaussian likelihood. For example in \cite{bib_non_gaussianity_shear_likelihood}, using independent component analysis, a significant non-Gaussianity in the likelihood is detected in the case of a cosmic shear study.

A more fundamental approach to this is performed in \cite{bib_peter_jan_paper}: Under very general conditions, it is shown that there exist constraints on the correlation function, meaning that a valid correlation function is confined to a certain range of values. This proves that the likelihood of the correlation function cannot be truly Gaussian: A Gaussian distribution has infinite support, meaning that it is non-zero also in ``forbidden'' regions.

Thus, it is necessary to look for a better description of the likelihood than a Gaussian. This could lead to a more accurate data analysis and thus, in the end, result in more precise constraints on the cosmological parameters obtained by such an analysis. It is also worth noting that \cite{bib_carron_2012} has shown some inconsistencies of Gaussian likelihoods: For the case of power spectrum estimators, using Gaussian likelihoods can assign too much information to the data and thus violate the Cram\'{e}r-Rao inequality. Along the same lines, \cite{bib_sun_ps_likelihood} show that the use of Gaussian likelihoods in power spectra analyses can have significant impact on the bias and uncertainty of estimated parameters.

Of course, non-Gaussian likelihood functions have been studied before: For the case of the cosmic shear power spectrum, \cite{bib_copula_likelihood} use a Gaussian copula to construct a more accurate likelihood function. The use of a copula requires knowledge of the univariate distribution, for which the authors find a good approximation from numerical simulations.

However, in the case of correlation functions, similar work is sparse in the literature, so up to now, no better general approximation than the Gaussian likelihood has been obtained. Obviously, the most fundamental approach to this is to try and calculate the probability distributions of correlation functions analytically: \cite{bib_david_paper} used a Fourier mode expansion of a Gaussian random field in order to obtain the characteristic function of $p(\xi)$. From it, they calculate the uni- and bivariate probability distribution of $\xi$ through a Fourier transform. Since this calculation turns out to be very tedious, an analytical computation of higher-variate PDFs is probably not feasible. Still, coupling the exact univariate distributions with a Gaussian copula might yield a multivariate PDF that is a far better approximation than the Gaussian one -- we will show, however, that this is not the case.

In the main part of this work, we will therefore try a different approach: Using a transformation of variables that involves the aforementioned constraints on correlation functions derived in \cite{bib_peter_jan_paper}, we will derive a way of computing a new, ``quasi-Gaussian'' likelihood function for the correlation functions of one-dimensional Gaussian random fields.

It is notable that the strategy of transforming a random variable in order to obtain a well-known probability distribution (usually a Gaussian one) suggests a comparison to similar attempts: For example, Box-Cox methods can be used to find an optimal variable transformation to Gaussianity (see \eg \citealt{bib_boxcox_benjamin}). We will argue, however, that a Box-Cox approach does not seem to yield satisfactory results in our case.

This work is structured as follows: In \autoref{sec:new_approximation}, we briefly summarize previous work and explain how to compute the quasi-Gaussian likelihood for correlation functions, which is then tested thoroughly and compared to numerical simulations in \autoref{sec:quality_of_qg}. In \autoref{sec:bayesian_analysis}, we investigate how the new approximation of the likelihood performs in a Bayesian analysis of simulated data, compared to the usual Gaussian approximation. Our attempts to find a new approximation of the likelihood using a copula or Box-Cox approach are presented in \autoref{sec:copula_boxcox}. We conclude with a short summary and outlook in \autoref{sec:conclusions}.

%__________________________________________________________________

\section{A new approximation for the likelihood of correlation functions}
\label{sec:new_approximation}

\subsection{Notation}
We denote a random field by $g(\vec{x})$ and define its Fourier transform as
\begin{equation}
 \tilde g(\vec{k})=\int\dd^n x\ g(\vec{x})\ \e^{-\im\vec{k}\cdot\vec{x}}.
 \label{eq:random_field_FT}
\end{equation}
In order to perform numerical simulations involving random fields, it is necessary to discretize the field, \ie evaluate it at discrete grid points, and introduce boundary conditions: Limiting the field to an $n$-dimensional periodic cube with side length $L$ in real space is equivalent to choosing a grid in $\vec{k}$-space such that the wave number can only take integer multiples of the smallest possible wave number $\Delta k=2\pi/L$. In order to simulate a field that spans the whole space $\Rn$, we can choose $L$ such that no structures at larger scales than $L$ exist. In this case, the hypercube is representative for the whole field and we can extend it to $\Rn$ by imposing periodic boundary conditions, \ie by setting (in one dimension) $g(x+L)=g(x)$. After discretization, the integral from \autoref{eq:random_field_FT} simply becomes a discrete Fourier series, and the field can be written as
\begin{equation}
 g(\vec{x})=\sum_{\vec{k}} \tilde g_{\vec{k}}\ \e^{\im\vec{k}\cdot\vec{x}},
\end{equation}
where we defined the Fourier coefficients
\begin{equation}
 \tilde g_{\vec{k}}=\left(\frac{\Delta k}{2\pi}\right)^n\tilde g(\vec{k}).
 \label{eq:g_fourier_coefficients}
\end{equation}
The two-point correlation function of the field is defined as $\xi(\vec x, \vec y)=\left<g(\vec x)\cdot g^*(\vec y)\right>;$ it is the Fourier transform of the power spectrum:
\begin{equation}
 \xi(|\vec x|) = \int\frac{\dd^n k}{(2\pi)^n}P(|\vec{k}|)\exp(\im\vec{k}\cdot\vec{x})
 = \int\frac{\dd k}{2\pi} k^{n-1}P(k)\ Z_n(kx),
 \label{eq:xi_FT}
\end{equation}
where the function $Z_n(\eta)$ is obtained from integrating the exponential over the direction of $\vec k$. In particular, $Z_2(\eta)=J_0(\eta)$ and $Z_3(\eta)=j_0(\eta)$, where $J_0(\eta)$ denotes the Bessel function of the first kind of zero order and $j_0(\eta)$ is the spherical Bessel function of zero order. In the one-dimensional case, \autoref{eq:xi_FT} becomes a simple cosine transform,
\begin{equation}
 \xi(x) = \int_0^\infty\frac{\dd k}{\pi}\ P(|k|)\cos(kx).
 \label{eq:xi_FT_1d}
\end{equation}
A random field is called Gaussian if its Fourier components $\tilde g_{\vec{k}}$ are statistically independent and the probability distribution of each $\tilde g_{\vec{k}}$ is Gaussian, \ie
\begin{equation}
 p(\tilde g_{\vec{k}})=\frac{1}{\pi\sigma^2(|\vec{k}|)}\exp\left(-\frac{\left|\tilde g_{\vec{k}}\right|^2}{\sigma^2(|\vec{k}|)}\right).
 \label{eq:prob_of_g_k}
\end{equation}
A central property of a Gaussian random field is that it is entirely specified by its power spectrum, which is given as a function of $\sigma(|\vec{k}|)$:
\begin{equation}
 P(|\vec{k}|)=\left(\frac{2\pi}{\Delta k}\right)^n\left<\left|\tilde g_{\vec{k}}\right|^2\right>= \left(\frac{2\pi}{\Delta k}\right)^n\sigma^2(|\vec{k}|).
 \label{eq:PS_sigma}
\end{equation}
In the following, we will only be concerned with one-dimensional, periodic Gaussian fields of field length $L$, evaluated at $N$ discrete grid points $g_i\equiv g(x_i)$, $x_i=i\ L/N$. 

\subsection{Constraints on correlation functions}
As shown in \cite{bib_peter_jan_paper} (hereafter \SH{}), correlation functions cannot take arbitrary values, but are subject to constraints, originating from the non-negativity of the power spectrum $P(k)$. The constraints are expressed in terms of the correlation coefficients
\begin{equation}
 r_n\equiv\frac{\xi(n\ \Delta x)}{\xi(0)}.
\end{equation}
Since we will be using a gridded approach, we shall denote $\xi(n\ \Delta x)\equiv\xi_n$, where $\Delta x=L/N$ denotes the separation between adjacent grid points. The constraints can be written in the form
\begin{equation}
 \rnl(r_1,r_2,\ldots,r_{n-1})\leq r_n\leq \rnu(r_1,r_2,\ldots,r_{n-1}),
\end{equation}
where the upper and lower boundaries are functions of the $r_i$ with $i<n$. \SH{} use two approaches to calculate the constraints: The first one applies the Cauchy-Schwartz inequality and yields the following constraints:
\begin{eqnarray}
 -1&\leq& r_1\leq 1,\\
 \label{eq:constraint_r1}
 -1+2r_1^2&\leq& r_2\leq 1,\\
 \label{eq:constraint_r2}
 -1+\frac{(r_1+r_{n-1})^2}{1+r_{n-2}}&\leq& r_n\leq 1-\frac{(r_1-r_{n-1})^2}{1-r_{n-2}},\quad n>2.
 \label{eq:constraints_from_cauchy}
\end{eqnarray}
The second approach involves the covariance matrix $C$, which for a one-dimensional random field $g_i$ of $N$ grid points, reads
\begin{equation}
 C_{ij}=\langle g_i g_j^* \rangle =\xi_{|i-j|}.
\end{equation}
Calculating the eigenvalues of $C$ and making use of the fact that they have to be non-negative due to the positive semi-definiteness of $C$ allows the calculation of constraints on $r_i$ for arbitrary $n$. As it turns out, for $n\geq 4$, the constraints obtained by this method are different (that is, stricter) than the ones from the Cauchy-Schwarz inequality -- \eg for $n=4$, while the upper bound remains unchanged compared to \autoref{eq:constraints_from_cauchy}, the lower bound then reads
\begin{equation}
 r_{4\mathrm{l}}=-1+\frac{r_1^2(1-4r_2)+2r_2^2(1+r_2)+2r_1 r_3(1-2r_2)+r_3^2}{1-2r_1^2+r_2}.
 \label{eq:constraint_r4_cov}
\end{equation}
Moreover, \SH{} show that the constraints are ``optimal'' (in the sense that no stricter bounds can be found for a general power spectrum) for the one-dimensional case. In more than one dimension, although the constraints derived by \SH{} still hold, stricter constraints will hold due to the multidimensional integration in \autoref{eq:xi_FT} and the isotropy of the field.

One can show that the constraints are obeyed by generating realizations of the correlation function of a Gaussian random field with a power spectrum that is randomly drawn for each realization (we will elaborate more on these simulations in \autoref{sec:simulations}).
\begin{figure}[!ht]
   \resizebox{0.95\hsize}{!}{\includegraphics[keepaspectratio, angle=-90]{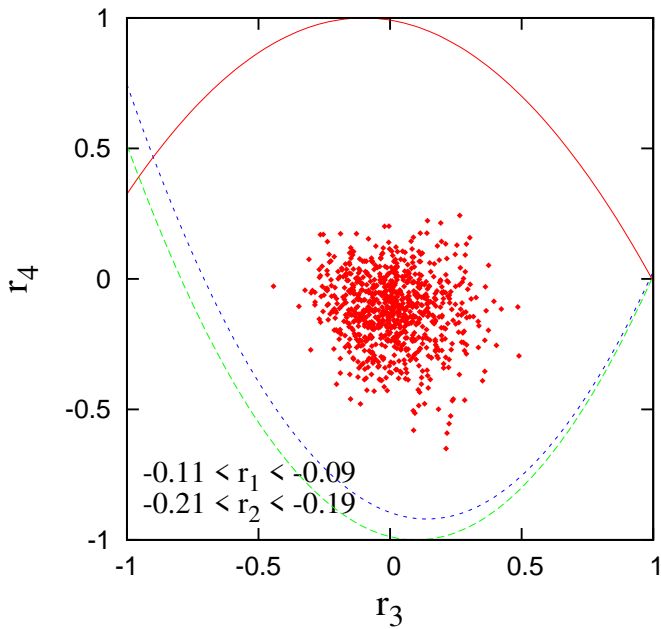}}\hfill
   \caption{Constraints in the $r_3$-$r_4$-plane (\ie with fixed $r_1$ and $r_2$) for a field with $N=16$ grid points and randomly drawn power spectra. For the lower bound, the results from the Cauchy-Schwarz inequality (green) as well as from the covariance matrix approach (blue) are shown.}
   \label{fig:scatter_r3_r4}
\end{figure}
In addition to the plots shown in \SH{}, we show the $r_3$-$r_4$-plane for illustration, see \autoref{fig:scatter_r3_r4}. Note that the admissible region is populated only sparsely, since the constraints on $r_3$ depend explicitly on $r_1$ and $r_2$, so we have to fix those, resulting in only 845 of the simulated $400\ 000$ realizations that have $r_1$ and $r_2$ close to the fixed values. \autoref{fig:scatter_r3_r4} shows the lower bounds $r_{4\mathrm l}$ as calculated from the Cauchy-Schwarz inequality and from the covariance matrix method. One can see that the latter (plotted in blue) is slightly stricter than the former, shown in green.

A central idea of \SH{} on the way to a better approximation for the PDF of correlation functions is to transform the correlation function to a space where no constraints exist, and where the Gaussian approximation is thus better justified. This can be done by defining
\begin{equation}
 y_n=\atanh\frac{2r_n-r_{n\mathrm u}-r_{n\mathrm l}}{r_{n\mathrm u}-r_{n\mathrm l}}\equiv\atanh\left(x_n\right).
 \label{eq:r_to_y}
\end{equation}
The transformation to $x_n$ maps the allowed range of $r_n$ to the interval $\left(-1,+1\right)$, which is then mapped onto the entire real axis $\left(-\infty,+\infty\right)$ by the inverse hyperbolic tangent. The choice of atanh by \SH{} was an ``educated guess'', rather than based on theoretical arguments.

\subsection{Simulations}
\label{sec:simulations}
In the following, we will explain the simulations we use to generate realizations of the correlation function of a Gaussian random field with a given power spectrum. A usual method to do this is the following: One can draw realizations of the field in Fourier space, \ie according to \autoref{eq:prob_of_g_k}, where the width of the Gaussian is given by the theoretical power spectrum via \autoref{eq:PS_sigma}. The field is then transformed to real space, and $\xi$ can then be calculated using an estimator based on the $g_i$.

We developed an alternative method: Since we are not interested in the field itself, we can save memory and CPU time by directly drawing realizations of the power spectrum. To do so, we obviously need the probability distribution of the power spectrum components $P_i\equiv P(k_i)\equiv P(i\ \Delta k)$ -- this can be calculated easily from \autoref{eq:prob_of_g_k}, since for one realization, $P_i=\left(2\pi/\Delta k\right)\left|\tilde g_i\right|^2=L\left|\tilde g_i\right|^2$. The calculation yields
\begin{equation}
 p(P_i)=\frac{1}{L\sigma_i^2}\exp\left(-\frac{P_i}{L\sigma_i^2}\right)\ H(P_i),
 \label{eq:prob_of_P_k}
\end{equation}
where $H(\xi)$ denotes the Heaviside step function. Thus, for Gaussian random fields, each power spectrum mode follows an exponential distribution with a width given by the model power $L\sigma_i^2$.

For each realization of the power spectrum, the correlation function can then be computed using a discrete cosine transform given by a discretization of \autoref{eq:xi_FT_1d},
\begin{equation}
 \xi_m=\frac{2}{L}\sum_{n=1}^{N/2}P_n\cos\frac{2\pi mn}{N}.
 \label{eq:xi_from_P}
\end{equation}
This method can in principle be generalized to higher-dimensional fields, where the savings in memory and CPU usage will of course be far greater. It is clear that this new method should yield the same results as the established one (which involves generating the field). We confirmed this fact numerically; additionally, we give a proof for the equivalence of the two methods in \autoref{sec:equivalence_sims}.

In order to draw the power spectrum components $P_i$ according to the PDF given in \autoref{eq:prob_of_P_k}, we integrate and invert this equation, yielding 
\begin{equation}
 P_i=-L\sigma_i^2\ln\left(u\right),
 \label{eq:P_k_from_uniform}
\end{equation}
where $u$ is a uniformly distributed random number with $u \in [0,1]$ (see \citealt{bib_nr} for a more elaborate explanation on random deviates).

As an additional improvement to our simulations, we implement a concept called quasi-random (or ``sub-random'') sampling; again, see \cite{bib_nr} for more details: Instead of drawing uniform deviates $u$ for each component $P_i$, we use quasi-random points, thus ensuring that we sample the whole hypercube of power spectra as fully as possible, independent of the sample size. One way to generate quasi-randomness is by using a so-called Sobol' sequence, see \eg \cite{bib_qrand_paper} for explanations. Implementations for Sobol' sequences exist in the literature, we use a C++ program available at the authors' website \url{http://web.maths.unsw.edu.au/~fkuo/sobol/}.
\begin{figure*}[!ht]
    \resizebox{0.49\hsize}{!}{\includegraphics[keepaspectratio, angle=-90]{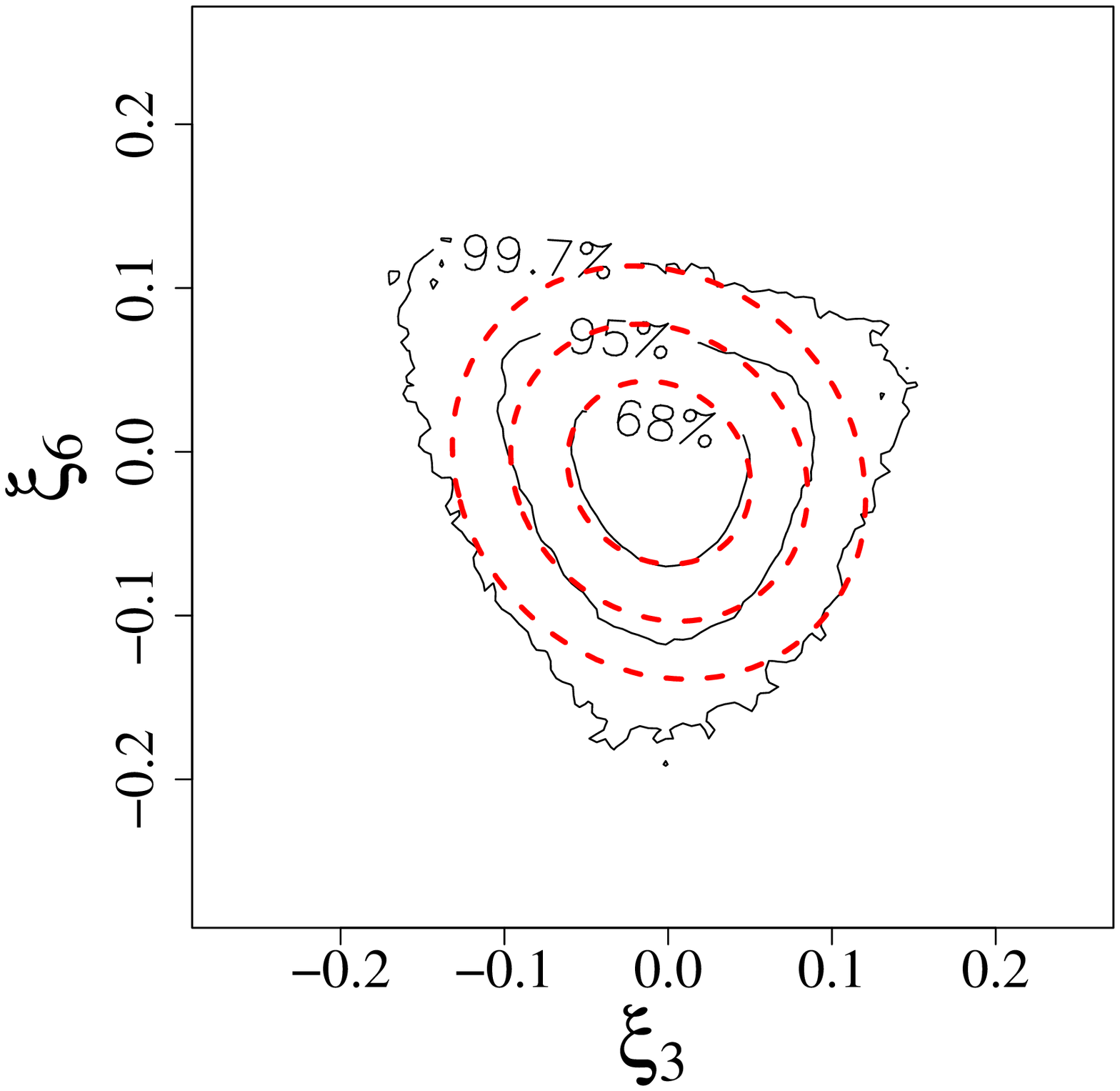}}\hfill
    \resizebox{0.48\hsize}{!}{\includegraphics[keepaspectratio, angle=-90]{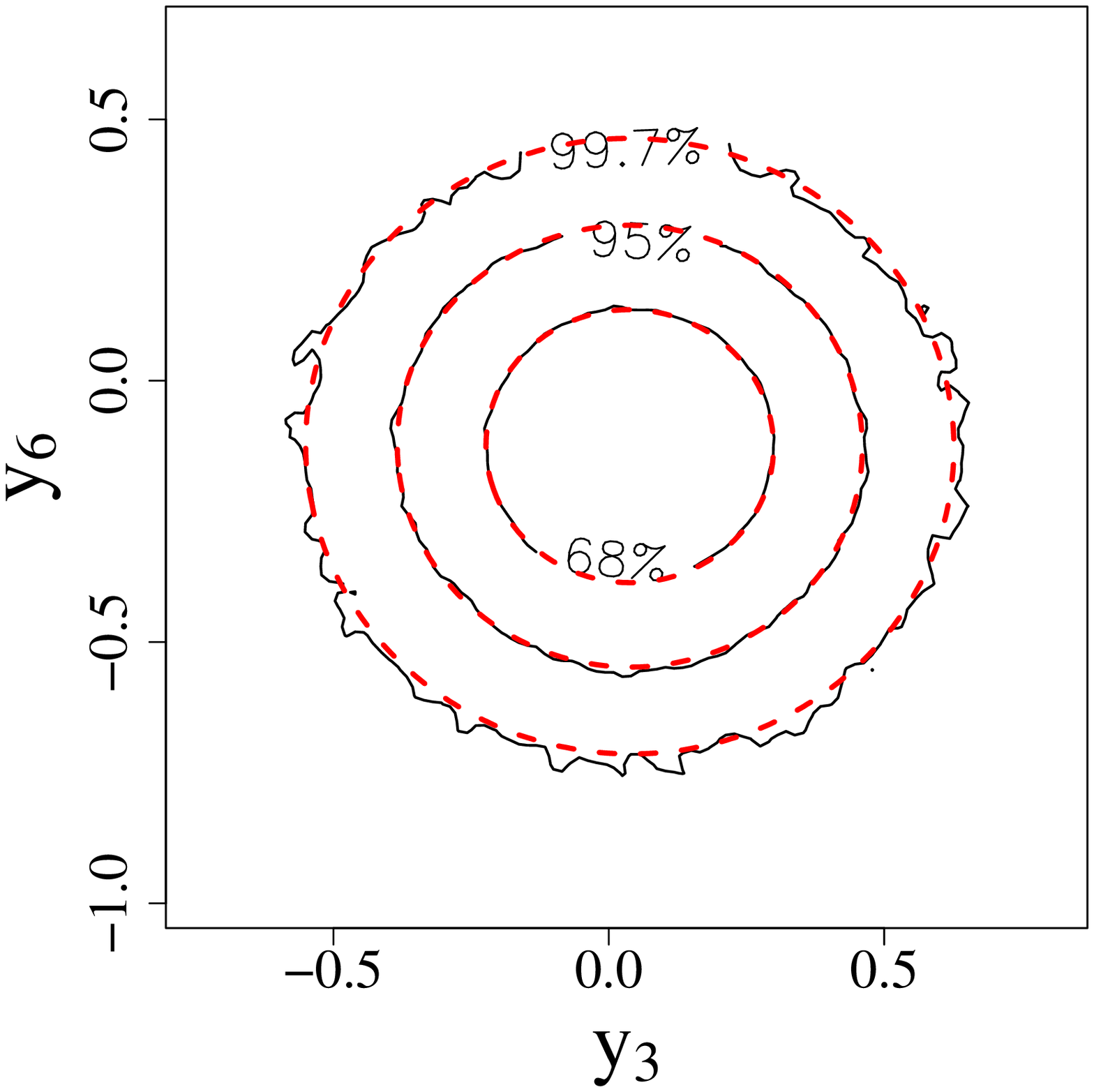}}
   \caption{Iso-probability contours of $p(\xi_3,\xi_6)$ (left) and $p(y_3,y_6)$ (right) for $400\ 000$ realizations of a field with $N=32$ grid points and a Gaussian power spectrum with $L k_0=80$. The dashed red contours show the best-fitting Gaussian approximations in both cases.}
   \label{fig:p_xi3xi6_y3y6_gauss}
\end{figure*}

To illustrate the non-Gaussian probability distribution of correlation functions, the left-hand part of \autoref{fig:p_xi3xi6_y3y6_gauss} shows the iso-probability contours of $p(\xi_3,\xi_6)$, \ie the PDF of the correlation function for a lag equal to 3 and 6 times the distance between adjacent grid points, calculated over the $400\ 000$ realizations of a field with $N=32$ grid points.
For the shape of the power spectrum, we choose a Gaussian, \ie $\sigma^2(k)\propto\exp\left(-\left(k/k_0\right)^2\right)$, whose width $k_0$ is related to the field length $L$ by $L k_0=80$. For simplicity, we will refer to this as a ``Gaussian power spectrum'' in the following, indicating its shape $P(k)$. The chosen Gaussian shape of the power spectrum should not be confused with the Gaussianity of the random field.
The red contours show the corresponding Gaussian with the same mean and standard deviation. It is clearly seen that the Gaussian approximation is not particularly good.
In order to check whether the Gaussian approximation is justified better in the new variables $y_i$, we transform the simulated realizations of the correlation function to $y$-space, using the constraints $\riu$ and $\ril$ from the covariance matrix approach, which we calculate exploiting the matrix methods explained in \SH{}. The right-hand part of \autoref{fig:p_xi3xi6_y3y6_gauss} shows the probability distribution $p(y_3,y_6)$ and the Gaussian with mean and covariance as computed from $\lbrace y_3,y_6\rbrace$. Obviously, the transformed variables follow a Gaussian distribution much more closely than the original correlation coefficients.

\subsection{Transformation of the PDF}
Before actually assessing how good the Gaussian approximation in $y$-space is, we will first try to make use of it and develop a method to obtain a better approximation of the probability distributions in $r$- and, in the end, $\xi$-space.

Assuming a Gaussian distribution in $y$-space, we can calculate the PDF in $r$-space as
\begin{equation}
 p_r\left(r_1,\dots,r_n\right)=p_y\left(y_1,\dots,y_n\right)\cdot\left|\det\left(J^{r\rightarrow y}\right)\right|.
\end{equation}
Here, $J^{r\rightarrow y}$ denotes the Jacobian matrix of the transformation defined in \autoref{eq:r_to_y}, so
\begin{eqnarray}
 J_{ij}^{r\rightarrow y}&=& \frac{\partial y_i}{\partial r_j}=\frac{\partial y_i}{\partial x_i}\frac{\partial x_i}{\partial r_j} %\nonumber\\
 =\frac{1}{1-x_i^2}\frac{\partial x_i}{\partial r_j}\nonumber\\
 &=& \frac{2}{\Deltai^2-(2r_i-\riu-\ril)^2}\nonumber\\
 &\times& \left[\Deltai\delta_{ij}-(r_i-\ril)\frac{\partial\riu}{\partial r_j}-(\riu-r_i)\frac{\partial\ril}{\partial r_j}\right].
 \label{eq:jacobian}
\end{eqnarray}
Since the lower and upper bounds $\ril$ and $\riu$ only depend on the $r_j$ with $j<i$, all partial derivatives with $i\leq j$ vanish. Thus, $J^{r\rightarrow y}$ is a lower triangular matrix, and its determinant can simply be calculated as the product of its diagonal entries. Hence, the partial derivatives $\partial\ril/\partial r_j$ and $\partial\riu/\partial r_j$ of the bounds are not needed.

\begin{figure}[!hb]
   \resizebox{0.95\hsize}{!}{\includegraphics[keepaspectratio, angle=-90]{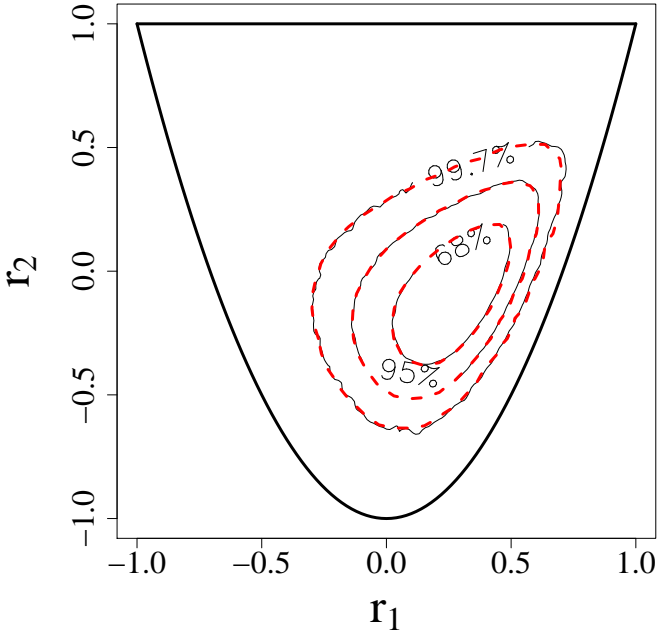}}\hfill
   \caption{$p(r_1,r_2)$ for a field with $N=32$ grid points and a Gaussian power spectrum with $L k_0=80$. The dashed red contours show the transformed Gaussian approximation from $y$-space.}
   \label{fig:p_r1r2_trafogauss}
\end{figure}

As an example, we calculate the bivariate distribution $p(r_1,r_2)$, assuming $p(y_1,y_2)$ to be a Gaussian with mean and covariance matrix obtained from the $400\ 000$ simulated realizations. \autoref{fig:p_r1r2_trafogauss} shows $p(r_1,r_2)$ from the simulations (solid contours) as well as the transformed Gaussian (dashed red contours) -- obviously, it is a far better approximation of the true distribution than a Gaussian in $r$-space. The upper and lower bounds on $r_2$ are also shown.

\begin{figure}[!ht]
    \resizebox{0.95\hsize}{!}{\includegraphics[keepaspectratio,angle=-90]{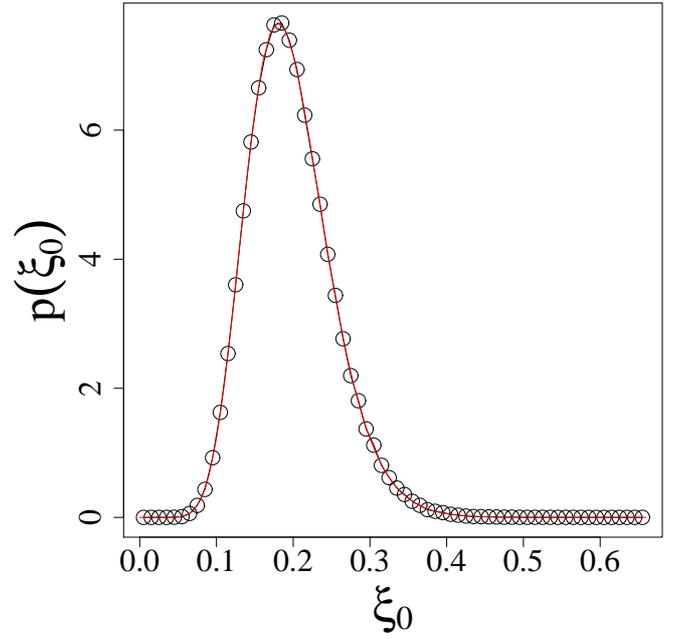}}
    \caption{$p(\xi_0)$ for a random field with $N=32$ grid points and a Gaussian power spectrum with $L k_0=80$: The black dots shows the PDF from the simulation, whereas the red curve was calculated from the analytical formula, using $N/2-1=15$ modes and the same field length and power spectrum.}
    \label{fig:p_xi0_analytic}
\end{figure}

In the next step, we will show how to compute the quasi-Gaussian likelihood for the correlation function $\xi$, since $\xi$ (and not $r$) is the quantity that is actually measured in reality.
The determinant of the Jacobian for the transformation $\xi\rightarrow r$ is simple, since $r_i=\xi_i/\xi_0$; however, we have to take into account that in $\xi$-space, an additional variable, namely $\xi_0$, exists. Thus, we have to write down the transformation as follows:
\begin{eqnarray}
 p(\xi_0,\xi_1,\dots,\xi_n)\prod_{i=0}^n\dd\xi_i &=& p'(\xi_1,\dots,\xi_n\mid\xi_0)\ p(\xi_0) \prod_{i=1}^n\dd\xi_i\nonumber\\
  &=& p_r(r_1,\dots,r_n\mid\xi_0)\ p(\xi_0) \prod_{i=1}^n\dd r_i\nonumber\\
  &=& p_y(y_1,\dots,y_n\mid\xi_0)\ p(\xi_0) \prod_{i=1}^n\dd y_i.
  \label{eq:p_trafo}
\end{eqnarray}
We see that the $n$-variate PDF in $\xi$-space explicitly depends on the distribution of the correlation function at zero lag, $p(\xi_0)$. In the following, we will use the analytical formula derived in \cite{bib_david_paper}. In the univariate case, it reads
\begin{eqnarray}
  p(\xi) &=& \sum_{n=1}^\infty \left\lbrace H(\xi)H(C_n)-H(-\xi)H(-C_n) \right\rbrace \nonumber\\
  &\times&\exp\left(-\frac{\xi}{2C_n}\right)\frac{1}{2C_n}\prod_{m\neq n}^\infty\frac{1}{1-\frac{C_m}{C_n}},
  \label{eq:p_xi_david}
\end{eqnarray}
where $H(\xi)$ again denotes the Heaviside step function and the $C_n$ are given by $C_n=\sigma_n^2\cos(k_n x)$. Here, $x$ is the lag parameter, and thus, for the zero-lag correlation function $\xi_0$, $C_n=\sigma_n^2$ holds. When evaluating \autoref{eq:p_xi_david}, we obviously have to truncate the summation at some $N'$. Since this number of modes $N'$ corresponds to the number of grid points in Fourier space (which is half the number of real-space grid points $N$, but as explained in \autoref{sec:equivalence_sims}, we set the highest mode to zero), we truncate at $N'=N/2-1$. As \autoref{fig:p_xi0_analytic} shows, the PDF of the $\lbrace\xi_0\rbrace$-sample (black dots) agrees perfectly with the result from the analytical formula (red curve).

As an illustrative example, we will show how to compute the quasi-Gaussian approximation of $p(\xi_1,\xi_2)$. In order to do so, we need to calculate the conditional probability $p(\xi_1,\xi_2|\xi_0)$ and then integrate over $\xi_0$. To perform the integration, we divide the $\xi_0$-range obtained from the simulations into bins and transform the integral into a discrete sum.
\begin{figure}[b]
   \resizebox{0.95\hsize}{!}{\includegraphics[keepaspectratio, angle=-90]{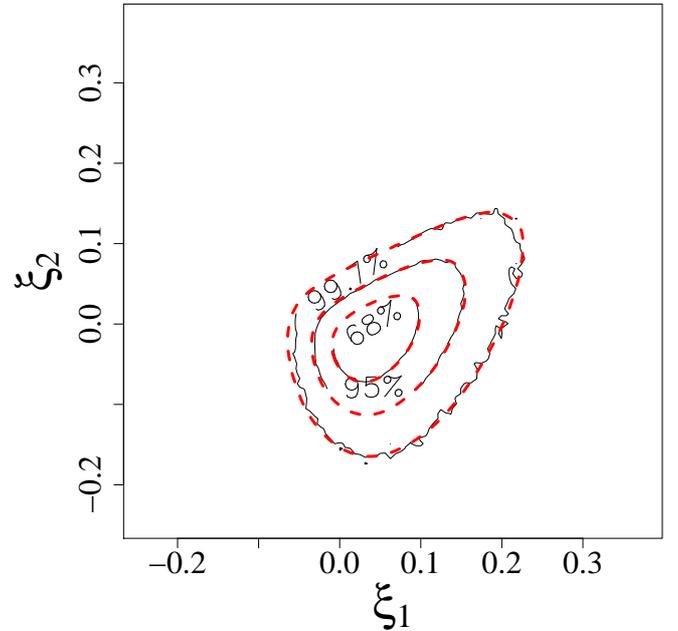}}\hfill
   \caption{$p(\xi_1,\xi_2)$ for a field with $N=32$ grid points and a Gaussian power spectrum with $L k_0=80$. The dashed red contours show the transformed Gaussian approximation from $y$-space.}
   \label{fig:p_xi1xi2_trafogauss}
\end{figure}

\begin{figure*}[ht!]
      \resizebox{0.95\hsize}{!}{\includegraphics[keepaspectratio, angle=-90]{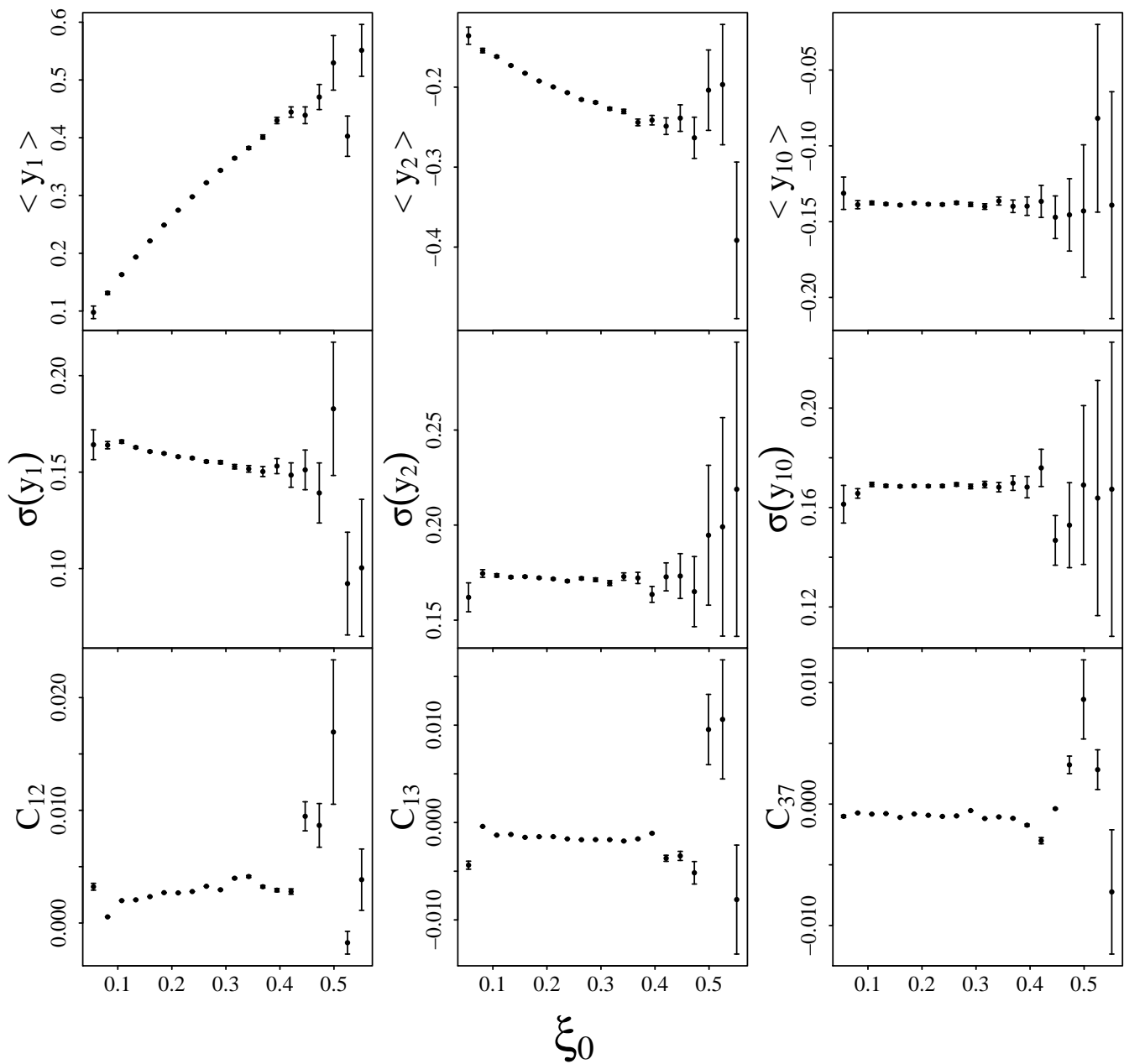}}
      \caption{The mean (top row) and standard deviation (second row) of $y_{n}$ for different $n$ as well as different off-diagonal elements of the covariance matrix $C\left(\langle y_{n}\rangle\right)$ (third row) as functions of $\xi_0$, determined from simulations; the error bars show the corresponding standard errors (which depend on the number of realizations with a value of $\xi_0$ that lies in the current bin).}
   \label{fig:xi0_dependence}
\end{figure*}

Furthermore, we have to incorporate a potential $\xi_0$-dependence of the covariance matrix and mean of $y$. We examine this dependence in \autoref{fig:xi0_dependence} -- clearly, the covariance matrix shows hardly any $\xi_0$-dependence, so as a first try, we treat it as independent of $\xi_0$ and simply use the covariance matrix obtained from the full sample. In contrast, the mean does show a non-negligible $\xi_0$-dependence, as the top panels of \autoref{fig:xi0_dependence} illustrate. However, this dependence seems to be almost linear, with a slope that decreases for higher lag parameter.
Thus, since the calculation of the multivariate quasi-Gaussian PDF involves computing a conditional probability with a fixed $\xi_0$-value as an intermediate step, we can easily handle the $\xi_0$-dependence by calculating the mean only over realizations close to the current value of $\xi_0$  -- in the exemplary calculation of $p(\xi_1,\xi_2)$ discussed here, we average only over those realizations with a $\xi_0$-value in the current bin of the $\xi_0$-integration.

Our final result for $p(\xi_1,\xi_2)$ is shown in \autoref{fig:p_xi1xi2_trafogauss}. Clearly, the quasi-Gaussian approximation (red dashed contours) follows the PDF obtained from the simulations quite closely. Note that the quasi-Gaussian PDF was calculated using only 10 $\xi_0$-bins for the integration, which seems sufficient to obtain a convergent result.

It should be noted that, although our phenomenological treatment of the $\xi_0$-dependence of mean and covariance matrix seems to be sufficiently accurate to give satisfying results, the analytical calculation of this dependence would of course be preferable, since it could improve the accuracy and the mathematical consistency of our method. We show this calculation in \autoref{sec:analytical_xi0_dep}.
However, since the results turn out to be computationally impractical and also require approximations, thus preventing a gain in accuracy, we refrain from using the analytical mean and covariance matrix.

%__________________________________________________________________

\section{Quality of the quasi-Gaussian approximation}
\label{sec:quality_of_qg}
\begin{figure*}[!ht]
   \resizebox{0.48\hsize}{!}{\includegraphics[keepaspectratio, angle=-90]{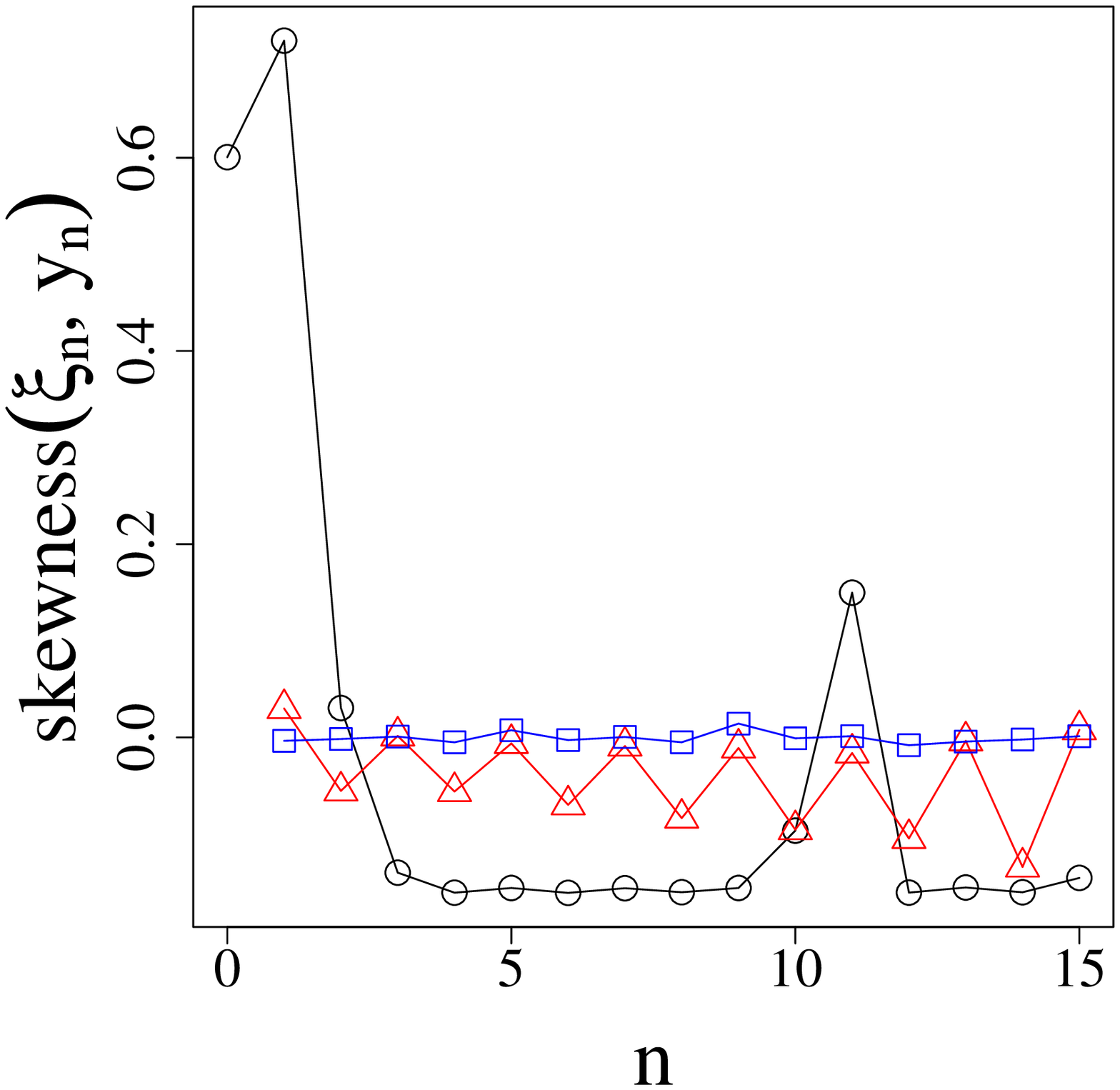}}\hfill
   \resizebox{0.48\hsize}{!}{\includegraphics[keepaspectratio, angle=-90]{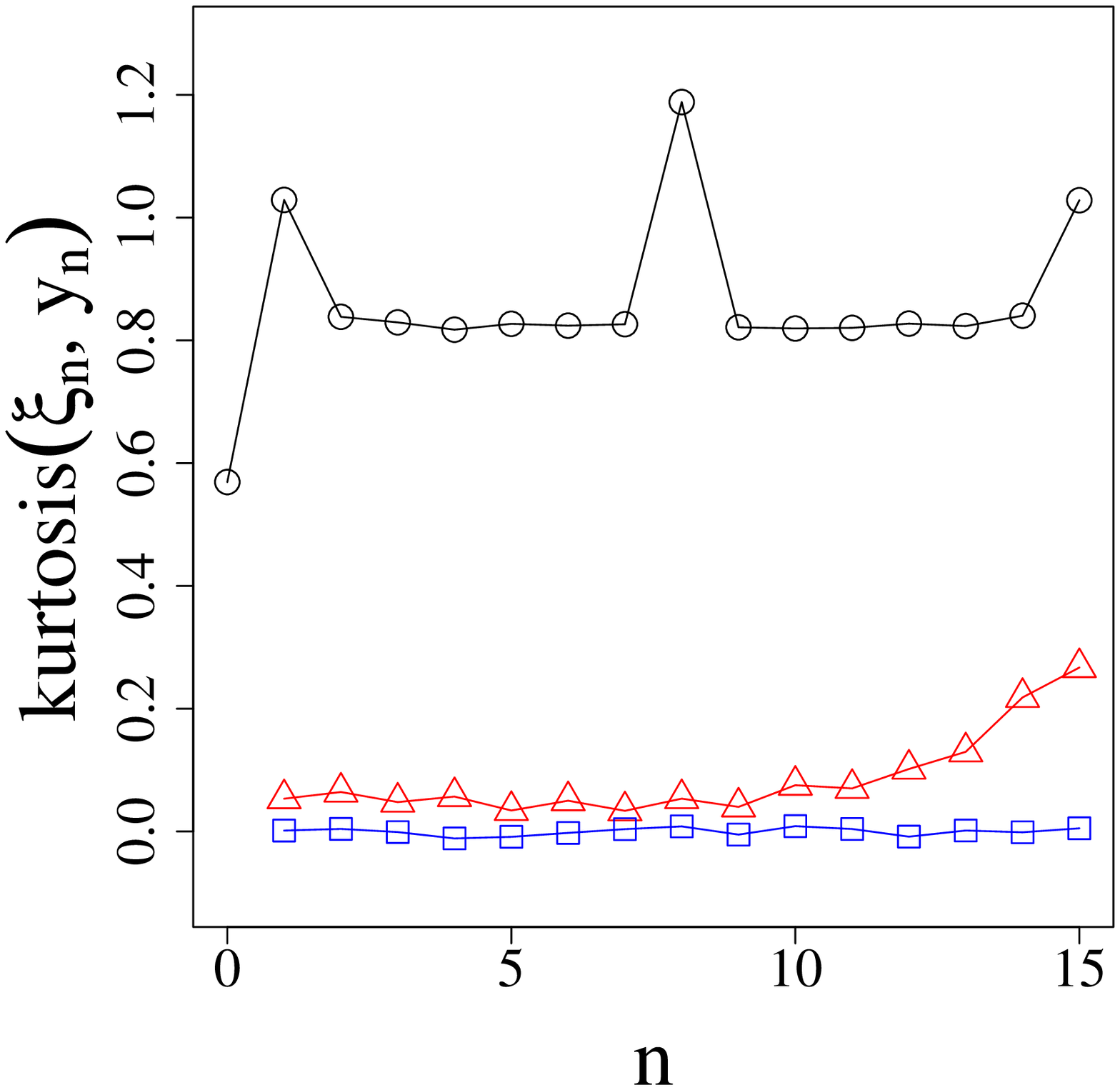}}
   \caption{Skewness (left-hand panel) and kurtosis (right-hand panel) of $\lbrace \xi_n\rbrace$ (black circles) and $\lbrace y_n\rbrace$ (red triangles) as functions of lag $n$, for a Gaussian power spectrum with $L k_0=80$. The blue squares show the skewness / kurtosis of Gaussian samples of the same length, mean and variance as the corresponding samples $\lbrace y_n\rbrace$.}
   \label{fig:skew_kurt_all}
\end{figure*}
For the bivariate distributions we presented so far, it is apparent at first glance that the quasi-Gaussian likelihood provides a more accurate approximation than the Gaussian one. However, we have not yet quantified how much better it actually is.

There are, in principle, two ways to address this: For one, we can directly test the Gaussian approximation in $y$-space. We will do so in \autoref{sec:quality_y}, making use of the fact that the moments of a Gaussian distribution are well-known. Alternatively, we can perform tests in $\xi$-space, which we will do in \autoref{sec:quality_xi}. Although in this case, we have to resort to more cumbersome methods, this approach is actually better, since it allows for a direct comparison to the Gaussian approximation in $\xi$-space. Furthermore, it is more significant, because it incorporates testing the transformation, especially our treatment of the $\xi_0$-dependence of the mean and covariance in $y$-space.

\subsection{Tests in \texorpdfstring{$y$}{y}-space}
\label{sec:quality_y}
In order to test the quality of the Gaussian approximation in $y$-space, we calculate moments of the univariate distributions $p(y_n)$, as obtained from our numerical simulations. The quantities we are interested in are the (renormalized) third-order moment, \ie the skewness
\begin{equation}
 \gamma=\left\langle\frac{(x-\mu)^3}{\sigma^3}\right\rangle\equiv\frac{m_3}{m_2^{3/2}}
 \label{eq:skewness}
\end{equation}
as well as kurtosis $\kappa$, which is essentially the fourth-order moment:
\begin{equation}
 \kappa=\left\langle\frac{(x-\mu)^4}{\sigma^4}\right\rangle-3\equiv\frac{m_4}{m_2^2}-3.
 \label{eq:kurtosis}
\end{equation}
Here, $m_i=\langle (x-\mu)^i\rangle$ denote the central moments. Defined like this, both quantities are zero for a Gaussian distribution. 

The results can be seen in \autoref{fig:skew_kurt_all}: The red triangles show the skewness / kurtosis of the $400\ 000$ simulated $\lbrace y_n\rbrace$-samples (again for a field with $N=32$ grid points and a Gaussian power spectrum with $L k_0=80$). Clearly, both moments deviate from zero, which cannot be explained solely by statistical scatter: The blue squares show the skewness / kurtosis of Gaussian samples of the same length, mean and variance as the corresponding samples $\lbrace y_n\rbrace$, and they are quite close to zero. However, it can also be clearly seen that the skewness and kurtosis of the univariate distributions in $\xi$-space (shown as black circles) deviate significantly more from zero, showing that the Gaussian approximation of the univariate distributions is far more accurate in $y$- than in $\xi$-space -- this is true also for other lag parameters than the ones used in the examples of the previous sections.

Note that although we used $N=32$ (real-space) grid points for our simulations, we only show the moments up to $n=15$, since the $\xi_n$ with higher $n$ contain no additional information due to the periodic boundary conditions (and thus, the corresponding $y_n$ have no physical meaning). Additionally, by construction, the $y_0$-component does not exist.

\begin{figure*}[!ht]
   \resizebox{0.48\hsize}{!}{\includegraphics[keepaspectratio, angle=-90]{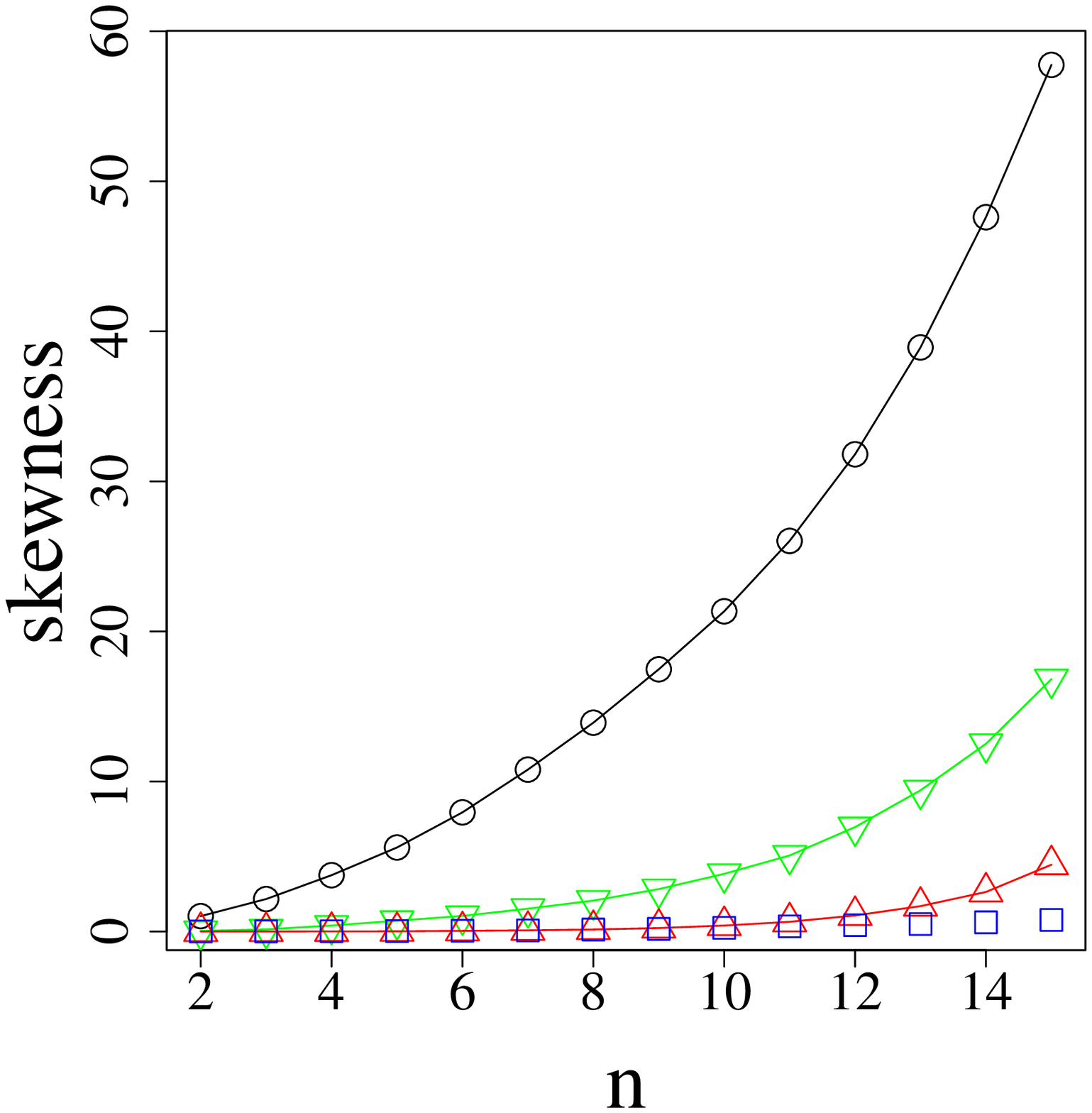}}\hfill
   \resizebox{0.48\hsize}{!}{\includegraphics[keepaspectratio, angle=-90]{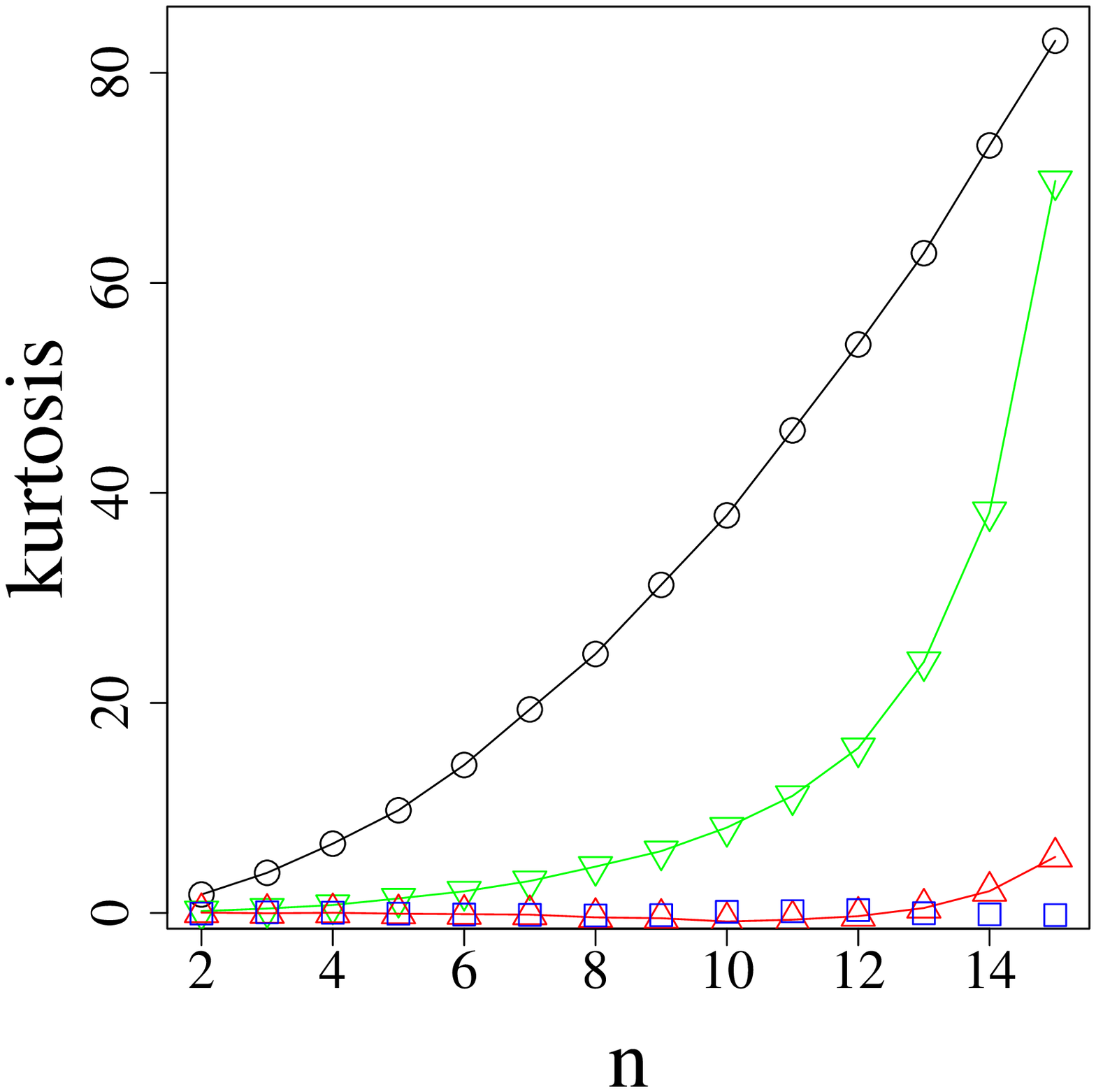}}
   \caption{Mardia's skewness (left-hand panel) and kurtosis (right-hand panel) of $n$-variate $\lbrace\xi\rbrace$- (black circles) and $\lbrace y\rbrace$-samples (red, upward triangles) for a Gaussian power spectrum with $L k_0=80$. The green, downward triangles show the skewness / kurtosis of $\lbrace y\rbrace$-samples obtained under the assumption of a Gaussian $\lbrace\xi\rbrace$-sample, and the blue squares show the skewness / kurtosis of Gaussian samples. See text for more details.}
   \label{fig:skew_kurt_mv_all}
\end{figure*}
Since the fact that the univariate PDFs $p(y_i)$ are ``quite Gaussian'' does not imply that the full multivariate distribution $p(\vec y)$ is well described by a multivariate Gaussian distribution, we also perform a multivariate test. While numerous tests for multivariate Gaussianity exist (see \eg \citealt{bib_mv_tests_review} for a review), we confine our analysis to moment-based tests. To do so, it is necessary to generalize skewness and kurtosis to higher dimensions -- we use the well-established definitions by \cite{bib_mardia_1970, bib_mardia_1974}: Considering a sample of $d$-dimensional vectors $x_i$, a measure of skewness for a $d$-variate distribution can be written as
\begin{equation}
 \gamma_d=\frac{1}{n^2}\sum_{i=1}^n\sum_{j=1}^n \left\lbrace \transpose{\left(\vec x_i -\vec\mu\right)} C^{-1} \left(\vec x_j -\vec\mu\right)\right\rbrace^3,
\end{equation}
where $n$ denotes the sample size, and $\vec\mu$ and $C$ are the mean and covariance matrix of the sample. The corresponding kurtosis measure is
\begin{equation}
 \kappa_d=\frac{1}{n}\sum_{i=1}^n \left\lbrace \transpose{\left(\vec x_i -\vec\mu\right)} C^{-1} \left(\vec x_i -\vec\mu\right)\right\rbrace^2 -d(d+2),
\end{equation}
where we subtracted the last term to make sure that the kurtosis of a Gaussian sample is zero.

\autoref{fig:skew_kurt_mv_all} shows the results of the multivariate test. For the same simulation run as before (using only 5000 realizations to speed up calculations), the skewness and kurtosis of the $n$-variate distributions, \ie $p(\xi_0,\ldots,\xi_{n-1})$ (black circles) and $p(y_1,\ldots,y_n)$ (red, upward triangles) are plotted as a function of $n$. For comparison, the blue squares show the moments of the $n$-variate distributions marginalized from a 15-variate Gaussian. It is apparent that also in the multivariate case, the assumption of Gaussianity is by far better justified for the transformed variables $y$ than for $\xi$, although the approximation is not perfect.

To avert comparing PDFs of different random variables (namely $\xi$ and $y$), we perform an additional check: We draw a 15-variate Gaussian sample in $\xi$-space and transform it to $y$-space (using only the realizations which lie inside the constraints); the corresponding values of skewness and kurtosis are shown as green, downward triangles. Clearly, they are by far higher than those obtained for the ``actual'' $y$-samples, further justifying our approach.                                                                                                                                                                                                                                                                                                                                                                            

\subsection{Tests in \texorpdfstring{$\xi$}{xi}-space}
\label{sec:quality_xi}
\begin{figure}[!hb]
   \resizebox{0.95\hsize}{!}{\includegraphics[keepaspectratio, angle=-90]{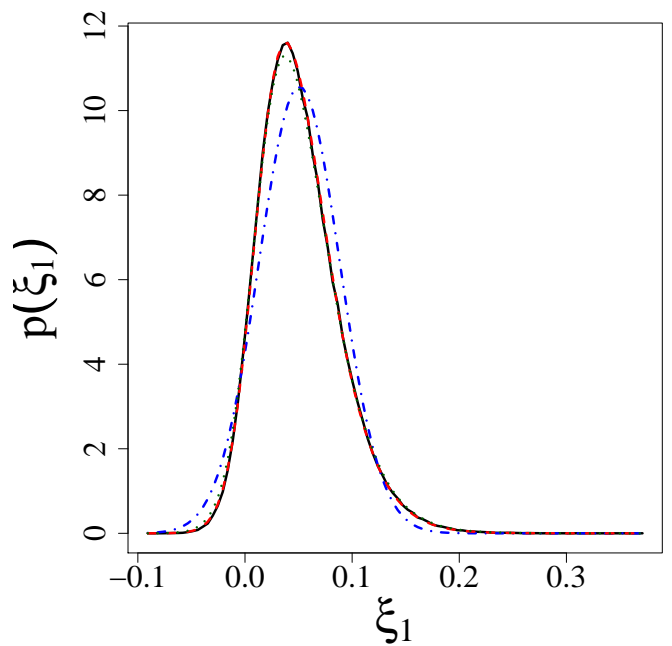}}\hfill
   \caption{$p(\xi_1)$ for a field with $N=32$ grid points and a Gaussian power spectrum with $L k_0=80$. The black curve is from simulations, the blue dot-dashed curve shows the best-fitting Gaussian, and the other two curves show the quasi-Gaussian approximation using a constant covariance matrix (green dotted curve) and incorporating its $\xi_0$-dependence (red dashed curve).}
   \label{fig:p_xi1_compare}
\end{figure}
In the following, we will directly compare the ``true'' distribution of the correlation functions as obtained from simulations to the Gaussian and the quasi-Gaussian PDF. As an example, we consider the univariate PDF $p(\xi_1)$. \autoref{fig:p_xi1_compare} shows the different distributions -- it is clear by eye that the quasi-Gaussian PDF is a far better approximation for the true (black) distribution than the best-fitting Gaussian (shown in blue and dot-dashed). In addition, it becomes clear that we can improve our results by incorporating the $\xi_0$-dependence of the covariance matrix: While the assumption of a constant covariance matrix already yields a quasi-Gaussian (shown as the green dotted curve) which is close to the true distribution, the red dashed curve representing the quasi-Gaussian which incorporates the $\xi_0$-dependence of the covariance matrix (in the same way as was done for the mean) is practically indistinguishable from the true PDF. In the following, we would like to quantify 
this result.

While a large variety of rigorous methods for comparing probability distributions exist, we will at first present an intuitive test: A straightforward way of quantifying how different two probability distributions $p(\xi)$ and $q(\xi)$ are is to integrate over the absolute value of their difference, defining the ``integrated difference''
\begin{equation}
 \Deltaint=\int\dd\xi\ \left|p(\xi)-q(\xi)\right|.
 \label{eq:Delta_integral}
\end{equation}
Here, we take $p$ to be the ``true'' PDF as obtained from the simulations, and $q$ the approximation. In order to calculate them, we introduce binning, meaning that we again split the range of $\xi_1$ in bins of width $\Delta\xi$ and discretize \autoref{eq:Delta_integral}. The values we obtain show a slight dependence on the number of bins -- the reason is Poisson noise, since $\Deltaint$ is not exactly zero even if the approximation is perfect, \ie if the sample with the empirical PDF $p$ was indeed drawn from the distribution $q$. The dependence on the binning becomes more pronounced in the multivariate case, but in the univariate example we consider here, $\Deltaint$ provides a coherent way of comparing the distributions.
Namely, if we divide the range of $\xi_1$ into 100 bins (the same number we used for \autoref{fig:p_xi1_compare}), we obtain $\Deltaint\approx 0.18$ for the Gaussian PDF and $\Deltaint\approx 0.025$ ($0.011$) for the quasi-Gaussian PDF with constant ($\xi_0$-dependent) covariance matrix, thus yielding a difference of more than one order of magnitude in the latter case.

A well-established measure for the ``distance'' between two probability distributions $p$ and $q$ is the Kullback-Leibler (K-L) divergence, which is defined as the difference of the cross-entropy $\mathcal H^X$ of $p$ and $q$ and the entropy $\mathcal H$ of $p$:
\begin{eqnarray}
 \DKL (p,q) =\mathcal H^X(p,q) - \mathcal H(p) =\int\dd \xi\  p(\xi) \log\frac{p(\xi)}{q(\xi)}.
 \label{eq:D_KL_integral}
\end{eqnarray}
As before, the values obtained vary slightly with the number of bins. Using 100 $\xi_1$-bins yields $\DKL\approx 0.04$ for the Gaussian and $\DKL\approx 0.001$ for quasi-Gaussian PDF with a constant covariance matrix. Here, incorporating the $\xi_0$-dependence of the covariance matrix has an even stronger impact, resulting in $\DKL\approx 0.0002$. Hence, the K-L divergence gives us an even stronger argument for the increased accuracy of the quasi-Gaussian approximation compared to the Gaussian case.

%______________________________________________________________

\section{Impact of the quasi-Gaussian likelihood on parameter estimation}
\label{sec:bayesian_analysis}
In this section, we will check which impact the new, quasi-Gaussian approximation of the likelihood has on the results of a Bayesian parameter estimation analysis. As data, we generate $400\ 000$ realizations of the correlation function of a Gaussian random field with $N=64$ grid points and a Gaussian power spectrum with $L\ k_0=100$.
From this data, we wish to obtain inference about the parameters of the power spectrum (\ie its amplitude $A$ and its width $k_0$) according to \autoref{eq:bayes}, so $\vec\theta=\left(A,k_0\right)$. 
To facilitate this choice of parameters, we parametrize the power spectrum as $P(k)=A\cdot 100/\left(L k_0\right)\ \exp\lbrace -\left(k/k_0\right)^2\rbrace$, where we choose $A=\unit[1]{Mpc}$ and $k_0=\unit[1]{Mpc^{-1}}$ as fiducial values (corresponding to $L\ k_0=100$ and a field length $L=\unit[100]{Mpc}$). Since we use a flat prior for $\theta$ and the denominator (the Bayesian evidence) acts solely as a normalization in the context of parameter estimation, the shape of the posterior $p(\vec{\theta}|\xi)$ is determined entirely by the likelihood. 

For the likelihood, we first use the Gaussian approximation
\begin{eqnarray}
 \mathcal{L}(\vec{\theta}) &\equiv& p(\xi|\vec{\theta}) = \frac{1}{\left(2\pi\right)^{n/2} \sqrt{\det C_\xi}} \nonumber\\
 &\times& \exp\left\lbrace -\frac{1}{2} \transpose{\left(\vec\xi\left(\vec\theta\right)-\vec\xi_\mathrm{fid}\right)} \cdot C_\xi^{-1}\cdot \left(\vec\xi\left(\vec\theta\right)-\vec\xi_\mathrm{fid}\right) \right\rbrace,
\end{eqnarray}
where $\vec\xi\equiv\transpose{\left(\xi_0,\dots,\xi_{n-1}\right)}$ and $C_\xi$ denotes the covariance matrix computed from the $\lbrace\vec\xi\rbrace$-sample; note that we use only $n=32$ $\vec\xi$-components, since the last 32 components yield no additional information due to periodicity.
In an analysis of actual data, the measured value of the correlation function would have to be inserted as mean of the Gaussian distribution -- since in our toy-model analysis, we use our simulated sample of correlation functions as data, we insert $\vec\xi_\mathrm{fid}\equiv\vec\xi\left(\vec\theta_\mathrm{fid}\right)$ instead, which is practically identical to the sample mean of the generated realizations.

For a second analysis, we transform the simulated realizations of the correlation function to $y$-space and adopt the quasi-Gaussian approximation, meaning that we calculate the likelihood as described in \autoref{sec:new_approximation}:
\begin{eqnarray}
 p(\xi|\vec{\theta}) &=& \frac{1}{\left(2\pi\right)^{(n-1)/2} \sqrt{\det C_y}} \nonumber\\
 &\times& \exp\left\lbrace -\frac{1}{2} \transpose{\left(\vec y\left(\vec\theta\right)-\left\langle\vec y\left(\xi_0\right)\right\rangle\right)} \cdot C^{-1}_y\cdot \left(\vec y\left(\vec\theta\right)-\left\langle\vec y\left(\xi_0\right)\right\rangle\right) \right\rbrace \nonumber\\
 &\times& p(\xi_0) \cdot \left|\det\left(J^{\xi\rightarrow y}\right)\right|.
 \label{eq:qg_likelihood}
\end{eqnarray}
Here, instead of inserting the fiducial value $\vec y_\mathrm{fid}$ as ``measured value'', we incorporate the $\xi_0$-dependence of the mean $\langle\vec y\rangle$ in the same way as previously described, meaning by calculating the average only over those realizations with a $\xi_0$-value close to the ``current'' value of $\xi_0$, which is the one determined by the fixed value of $\vec\theta$, \ie $\xi_0(\vec\theta)$.
In contrast to that, $C_y$ denotes the covariance matrix of the full $y$-sample, meaning that we neglect its $\xi_0$-dependence, although we have shown in \autoref{sec:quality_xi} that incorporating this dependence increases the accuracy of the quasi-Gaussian approximation. The reason for this is that for some values of $\xi_0$, the number of realizations with a $\xi_0$-value close to it is so small that the sample covariance matrix becomes singular.
However, since the toy-model analysis presented in this section is about a proof of concept rather than about maximizing the accuracy, this is a minor caveat -- of course, when applying our method in an analysis of real data, the $\xi_0$-dependence of $C_y$ should be taken into account. It should also be mentioned that apart from the $\xi_0$-dependence, we also neglect any possible dependence of the covariance matrix on the model parameters $\vec\theta$, since this is not expected to have a strong influence on parameter estimation (for example in the case of BAO studies, \citealt{bib_model_dep_cov_labatie_2012} show that even if $C$ does have a slight dependence on the model parameters, incorporating it in a Bayesian analysis only has a marginal effect on cosmological parameter constraints).

The $\vec\theta$-dependence of the last two terms also merits some discussion, in particular, the role of the fiducial model parameters $\vec\theta_\mathrm{fid}$ has to be specified: While the Gaussian likelihood discussed previously is of course centered around the fiducial values by construction, since $\vec\xi\left(\vec\theta_\mathrm{fid}\right)$ is inserted as mean of the Gaussian distribution, this cannot be done in the case of the quasi-Gaussian likelihood due to its more complicated mathematical form.

The $p(\xi_0)$-term in \autoref{eq:qg_likelihood} can be treated in a straightforward way: Namely, we fix the shape of the PDF $p(\xi_0)$ by determining it from the fiducial power spectrum parameters $\vec\theta_\mathrm{fid}$ and then evaluate it at the current value $\xi_0(\vec\theta)$. Thus, we compute this term as $p_{\vec\theta_\mathrm{fid}}(\xi_0(\vec\theta))$ -- in other words, we use $\vec\theta_\mathrm{fid}$ to fix $C_n$ as well as $C_m$ and then evaluate \autoref{eq:p_xi_david} at $\xi_0(\vec\theta)$.

The last term, \ie the determinant of the transformation matrix, however, has to be evaluated for the fiducial value, yielding $|\det(J^{\xi\rightarrow y}(\vec\theta_\mathrm{fid}))|$. Thus, this term has no $\vec\theta$-dependence at all and plays no role in parameter estimation; similarly to the Bayesian evidence, it does not have to be computed, but can rather be understood as part of the normalization of the posterior. This can be explained in a more pragmatic way: Assume that a specific value of $\vec\xi$ has been measured and one wants to use it for inference on the underlying power spectrum parameters, incorporating the quasi-Gaussian likelihood. Then one would transform the measurement to $y$-space and rather use the resulting $\vec y$-vector as data in the Bayesian analysis than $\vec\xi$, and thus, the $|\det J|$-term would not even show up when writing down the likelihood.
Here, we nonetheless included it in order to follow the train of thought from \autoref{sec:new_approximation} and for better comparison with the Gaussian likelihood in $\xi$-space.

Note that the previous argument cannot be applied to the $p(\xi_0)$-term: Calculating it as $p(\xi_0(\vec\theta_\mathrm{fid})|\vec\theta_\mathrm{fid})$ (\ie as independent of $\vec\theta$, making it redundant for parameter estimation), would yield biased results for two reasons: First, the transformation from $\vec\xi$ to $\vec y$ involves only ratios $\xi_i/\xi_0$, which means that one would neglect large parts of the information contained in $\xi_0$ -- in the case discussed here, this would lead to a large degeneracy in the amplitude $A$ of the power spectrum.
Furthermore, incorporating the non-negligible $\xi_0$-dependence of the mean $\langle\vec y\rangle$ immediately requires the introduction of the conditional probability in \autoref{eq:p_trafo}, thus automatically introducing the $p(\xi_0)$-term.
\begin{figure*}[!ht]
   \resizebox{0.48\hsize}{!}{\includegraphics[keepaspectratio, angle=-90]{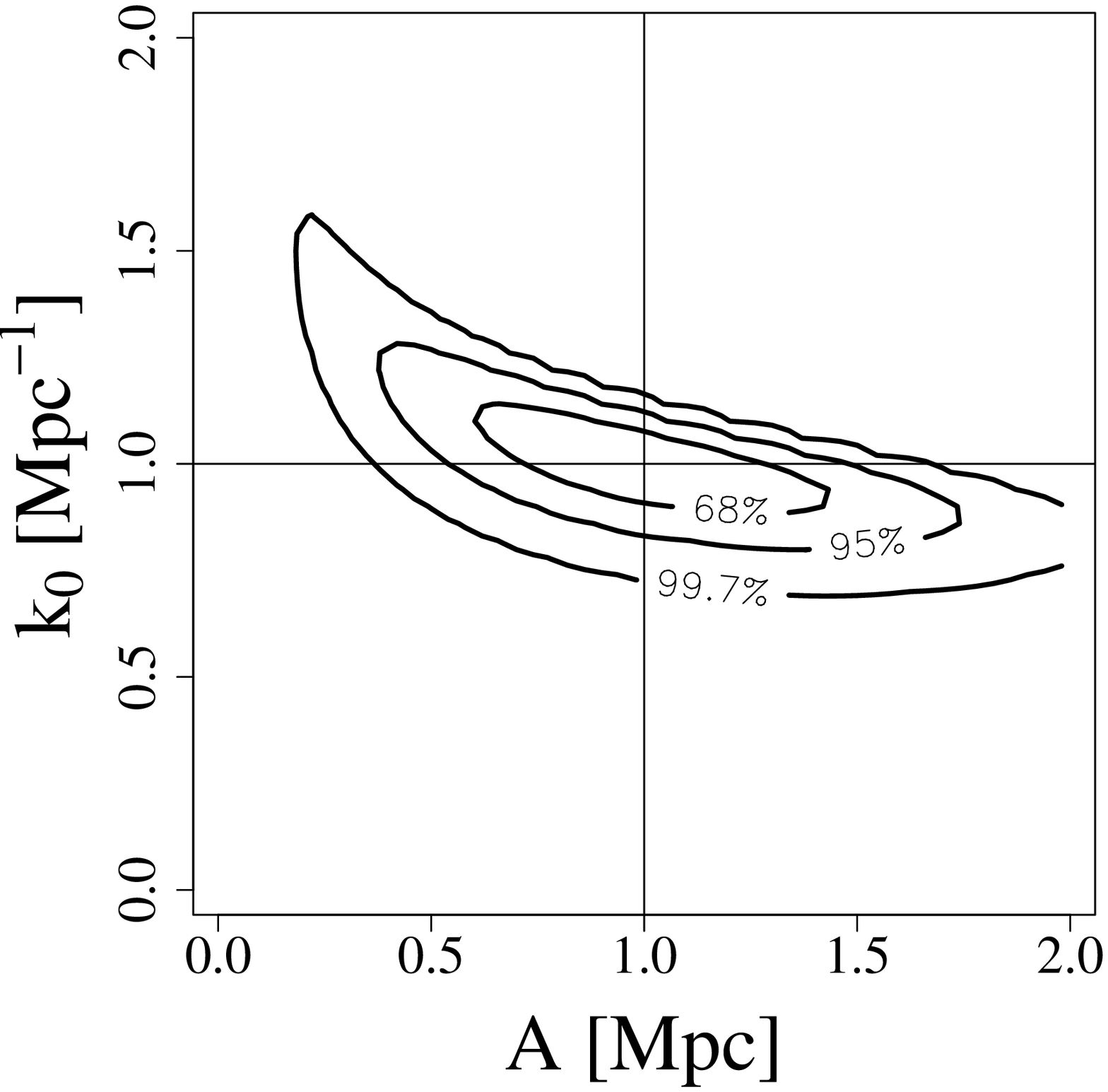}} \hfill
   \resizebox{0.48\hsize}{!}{\includegraphics[keepaspectratio, angle=-90]{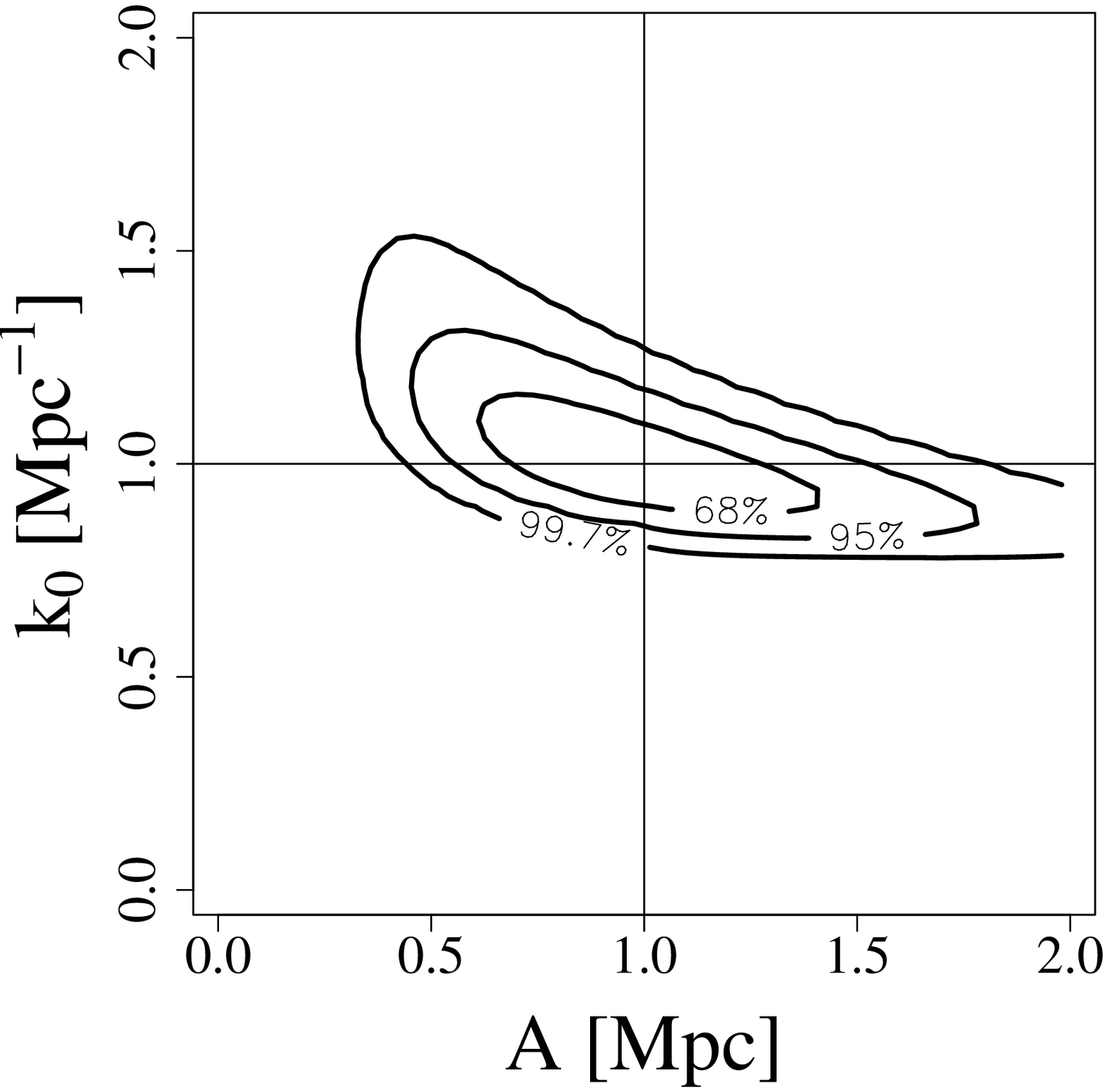}}
   \caption{Posterior probability for the power spectrum parameters $\left(A,k_0\right)$ using the Gaussian (left panel) and the quasi-Gaussian (right panel) likelihood. The horizontal and vertical lines are the fiducial values $\left(1.0,1.0\right)$.}
   \label{fig:posteriors}
\end{figure*}
\begin{figure*}[!ht]
   \resizebox{0.48\hsize}{!}{\includegraphics[keepaspectratio, angle=-90]{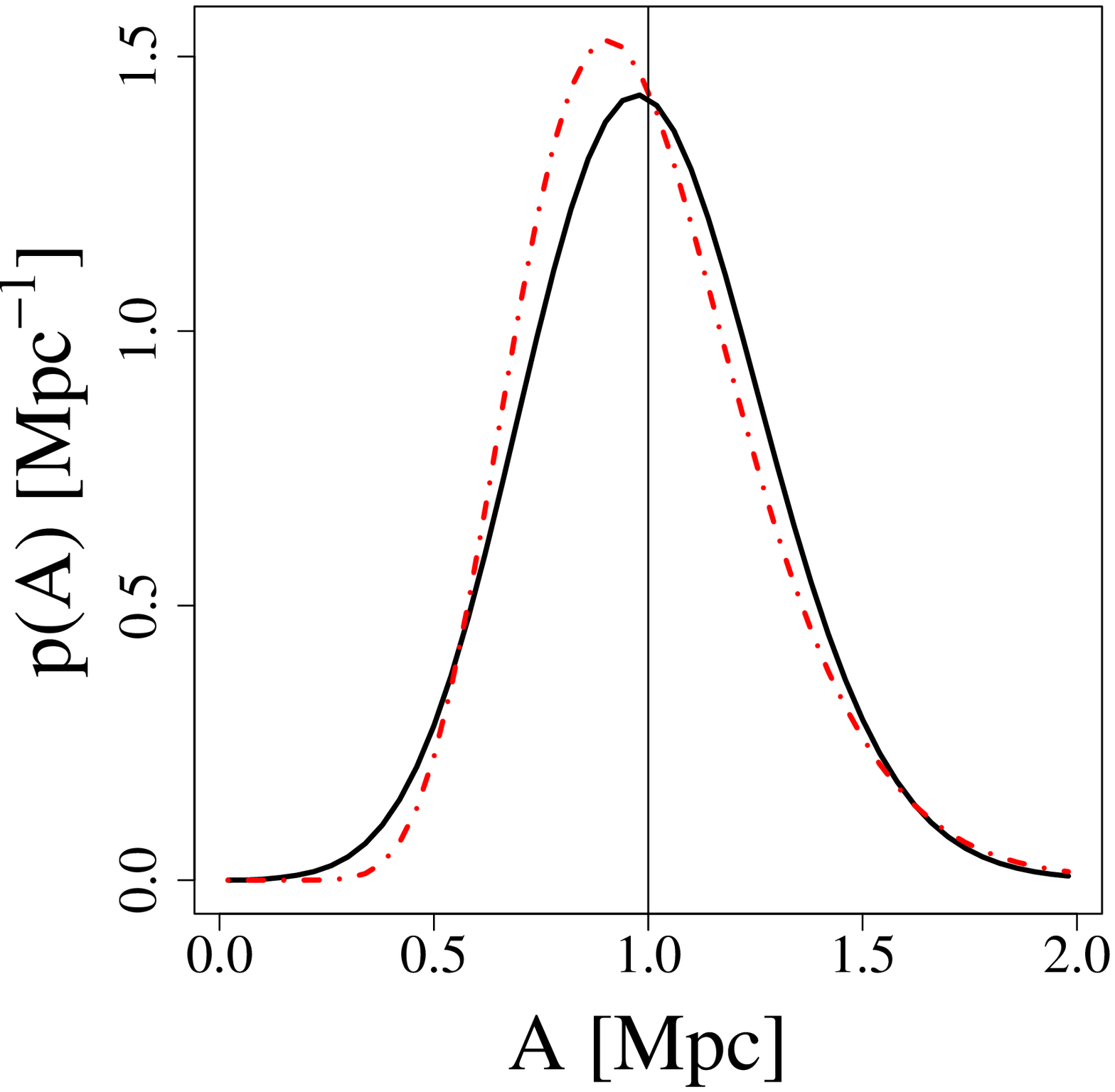}} \hfill
   \resizebox{0.48\hsize}{!}{\includegraphics[keepaspectratio, angle=-90]{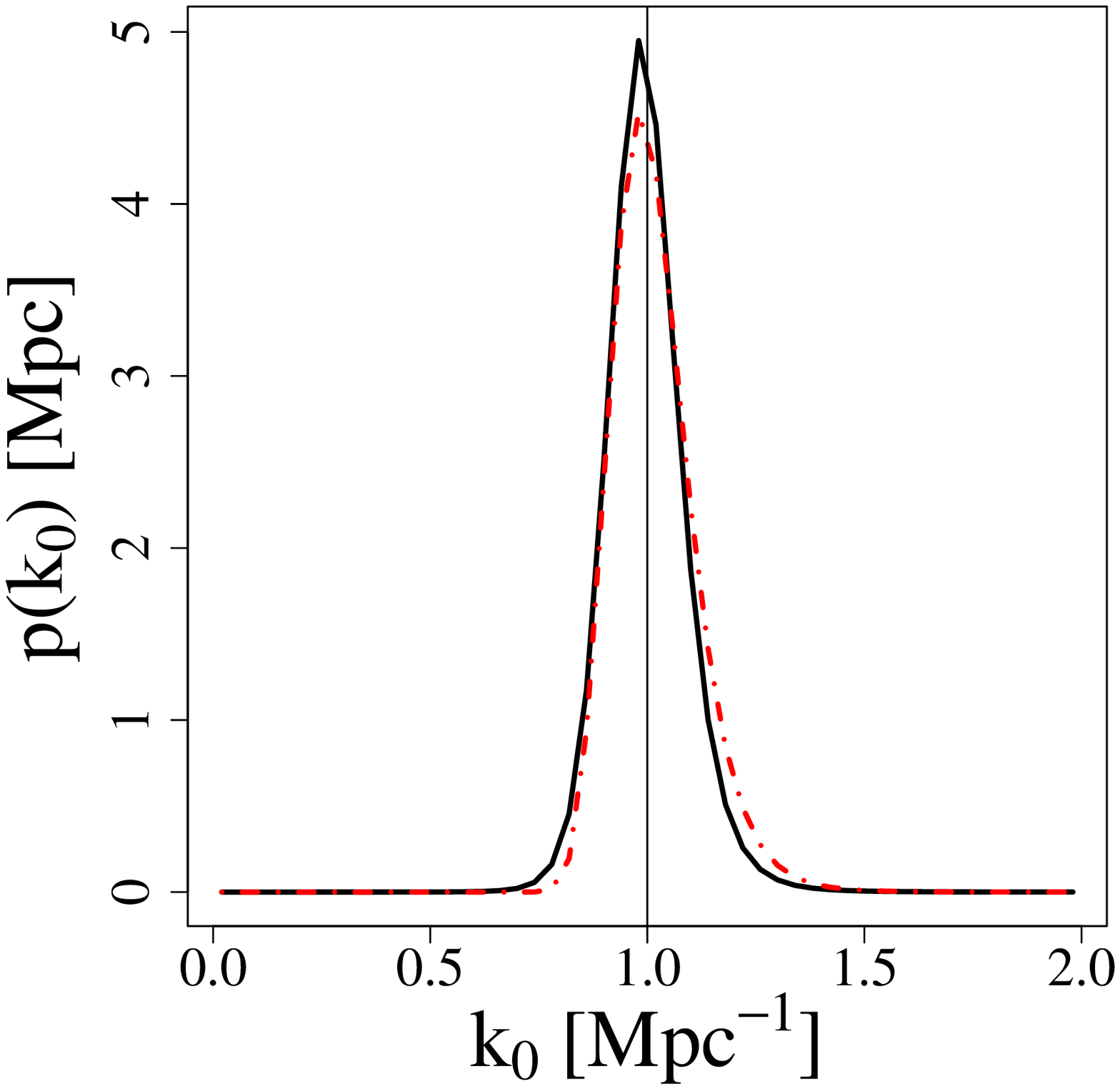}}
   \caption{Marginalized posterior probabilities for the power spectrum parameters $A$ and $k_0$. The solid black curves are the results obtained from the Gaussian likelihood, whereas the red dot-dashed curves show the results from the quasi-Gaussian analysis.}
   \label{fig:marg_posteriors}
\end{figure*}

The resulting posteriors can be seen in \autoref{fig:posteriors}, where the left-hand panel shows the result for the case of a Gaussian likelihood, and the right-hand one is the result of the quasi-Gaussian analysis. Already for this simple toy model, the impact of the more accurate likelihood on the posterior is visible. The difference may be larger for a different choice of power spectrum, where the deviation of the likelihood from a Gaussian is more pronounced. Nonetheless, it is evident that the change in the shape of the contours is noticeable enough to have an impact on cosmological parameter estimation.

\autoref{fig:marg_posteriors} shows the marginalized posteriors for $A$ and $k_0$, again for the Gaussian (black solid curve) and the quasi-Gaussian (red dot-dashed curve) case. As for the full posterior, there is a notable difference. While it may seem alarming that the marginalized posteriors in the quasi-Gaussian case are not centered around the fiducial value, this is in fact not worrisome: First, as a general remark about Bayesian statistics, it is important to keep in mind that there are different ways of obtaining parameter estimates, and in the case of a skewed posterior (as the quasi-Gaussian one), the ``maximum a posteriori'' (MAP) is not necessarily the most reasonable one. Furthermore, in our case, it should again be stressed that the Gaussian likelihood is of course constructed to be centered around the fiducial value, since $\vec\theta_\mathrm{fid}$ is explicitly put in as mean of the distribution, whereas the quasi-Gaussian likelihood is mainly constructed to obey the boundaries of the 
correlation functions. These are mathematically fundamental constraints, and thus, although none of the two methods are guaranteed to be bias-free, the quasi-Gaussian one should be favored.

%______________________________________________________________

\section{Alternative approaches}
\label{sec:copula_boxcox}
In this section, we briefly investigate two potential alternative ways of finding a new approximation for the likelihood of correlation functions, namely a copula approach and a method involving Box-Cox transformations.

\subsection{A copula approach}
\begin{figure*}[ht!]
    \resizebox{0.48\hsize}{!}{\includegraphics[keepaspectratio, angle=-90]{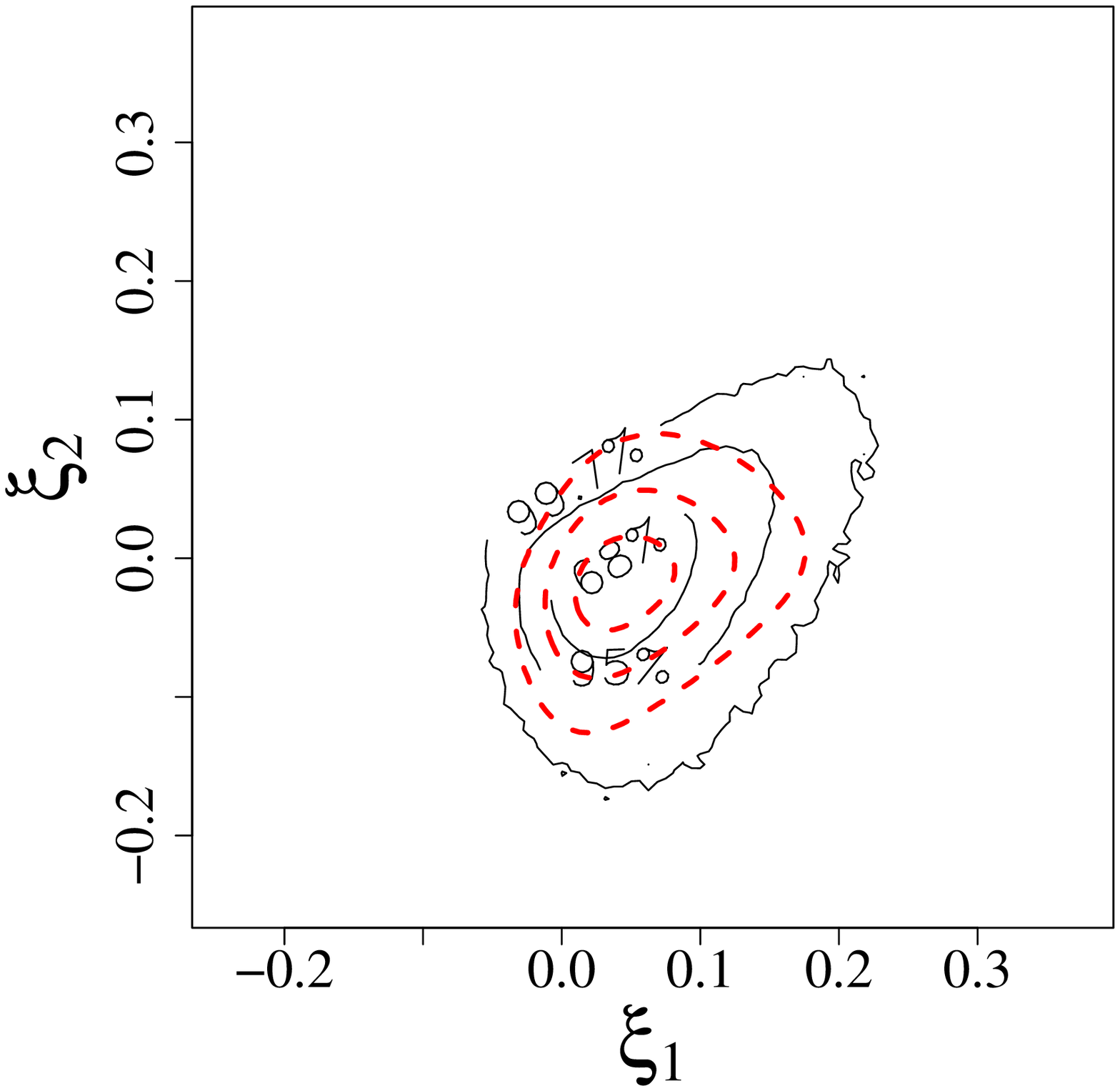}} \hfill
    \resizebox{0.48\hsize}{!}{\includegraphics[keepaspectratio, angle=-90]{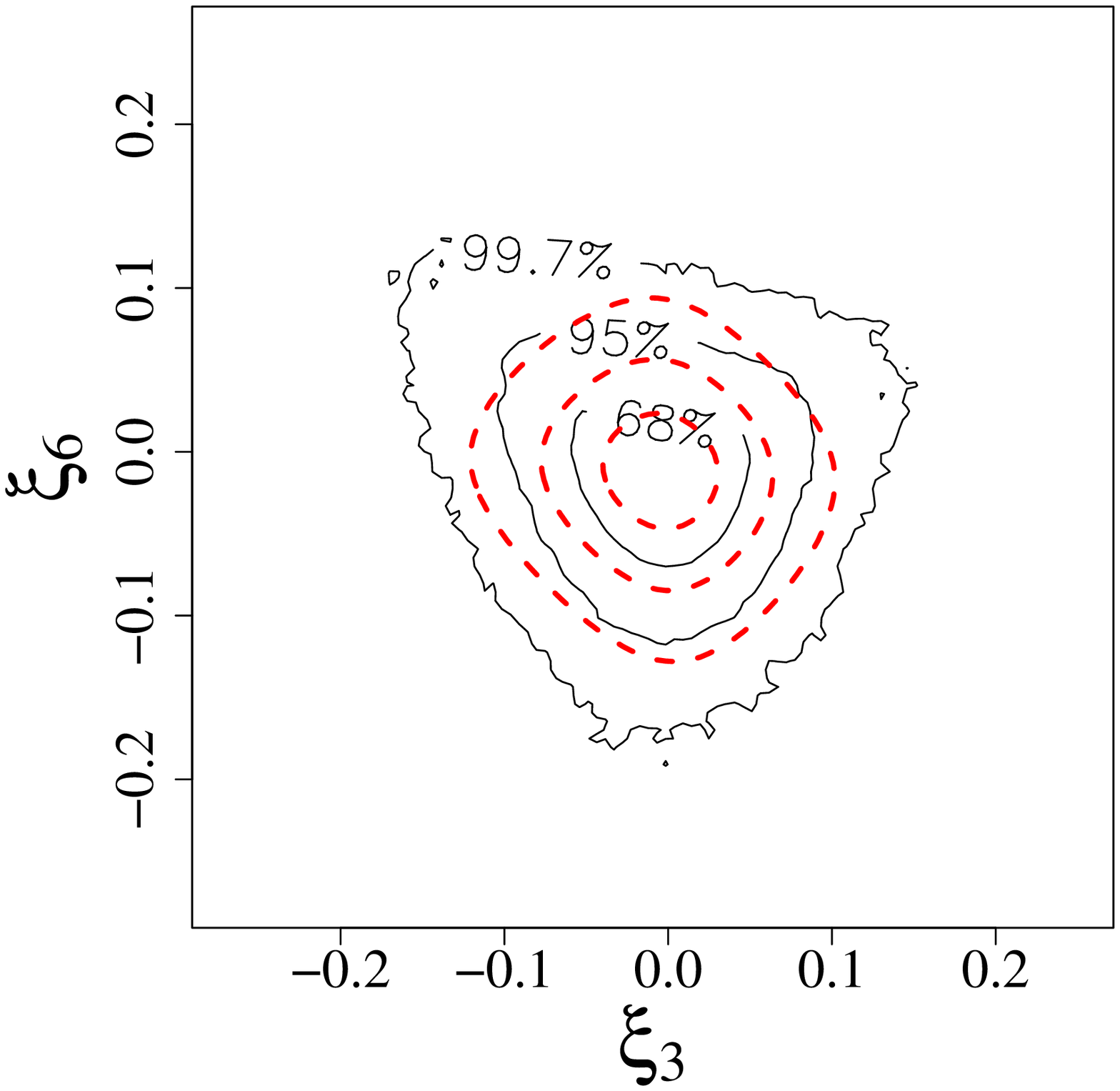}}
   \caption{$p(\xi_1,\xi_2)$ (left) and $p(\xi_3,\xi_6)$ (right) for a field with $N=32$ grid points and a Gaussian power spectrum with $L k_0=80$. The dashed red contours show the approximation obtained from a Gaussian copula.}
   \label{fig:copula}
\end{figure*}
Since the exact univariate PDF of $\xi$ is known (\citealt{bib_david_paper}), using a copula approach to compute the correlation function likelihood seems to be an obvious step. According to the definition of the copula, the joint PDF of $n$ random variables $\xi_i$ can be calculated from the univariate distributions $p_i(\xi_i)$ as
\begin{equation}
  p(\xi_1,\xi_2,\dots,\xi_n) = c(u_1,u_2,\dots,u_n)\cdot \prod_{i=1}^n p_i(\xi_i),
\end{equation}
where the copula density function $c$  depends on $u_i = P_i(\xi_i)$, \ie on the cumulative distribution functions (CDFs) of $\xi_i$. In the simplest case, the copula function is assumed to be a Gaussian.

Copulae have previously been used in cosmology, \eg for the PDF of the density field of large-scale structure (\citealt{bib_copula_scherrer}), or the weak lensing convergence power spectrum (see \citealt{bib_copula_likelihood} and its companion paper \citealt{bib_copula_cosm_parameters}). In the case of a Gaussian copula, the bivariate joint PDF can be calculated (see \eg \citealt{bib_copula_likelihood}) as
\begin{eqnarray}
 p\left(\xi_1,\xi_2\right) &=& \frac{1}{\sqrt{2\pi\ \det\left(C\right)}}\ \exp\left\lbrace -\frac{1}{2}\transpose{\left(\vec q -\vec \mu\right)} C^{-1} \left(\vec q -\vec \mu\right)\right\rbrace\nonumber\\
 &\times& \prod_{i=1}^2\left(\frac{1}{\sqrt{2\pi}\sigma_i}\ \exp\left\lbrace -\frac{\left(q_i-\mu_i\right)^2}{2\sigma_i^2}\right\rbrace  \right)^{-1}\cdot p\left(\xi_i\right),
 \label{eq:copula_xi_bivariate}
\end{eqnarray}
where $q_i = \Phi^{-1}_{\mu_i,\sigma_i}\left(P\left(\xi_i\right)\right)$. Here, $\vec\mu$ and $C$ denote the mean and covariance matrix of the copula function, $\Phi_{\mu,\sigma}$ is the Gaussian CDF. Note that contrary to our usual notation, here the indices of $\xi$ are purely a numbering, and \autoref{eq:copula_xi_bivariate} can in fact be applied for arbitrary lags.

To calculate the copula likelihood, we implement the analytical univariate formulae for $p_i(\xi_i)$ and $P_i(\xi_i)$ derived by \cite{bib_david_paper}; the mean and covariance matrix of the Gaussian copula are calculated directly from the simulated $\lbrace\vec\xi\rbrace$-sample.
\autoref{fig:copula} shows the bivariate PDFs from the simulation (black contours) as well as the copula likelihood (red dashed contours) for two different combinations of lags. It is apparent that the copula likelihood does not describe the true PDF very well. In particular, it does not even seem to be a more accurate description than the simple multivariate Gaussian used in the left-hand panel of \autoref{fig:p_xi3xi6_y3y6_gauss}, leading us to the conclusion that our quasi-Gaussian approximation should be favored also over the copula likelihood. Of course, the accuracy of the latter might improve if a more realistic coupling than the Gaussian one was found.

%______________________________________________________________

\subsection{Box-Cox transformations}
As mentioned in \autoref{sec:intro}, the idea of transforming a random variable in order to Gaussianize it suggests testing the performance of Box-Cox transformations. They are a form of power transforms originally introduced by \cite{bib_boxcox_orig} and have been used in astronomical applications, see \eg \cite{bib_boxcox_benjamin} for results on Gaussianizing the one-point distributions of the weak gravitational lensing convergence.

For a sample of correlation functions $\lbrace\xi_i\rbrace$ at a certain lag, the form of the Box-Cox transformation is
\begin{equation}
 \bar{\xi}_i(\lambda_i,a_i) = \left\lbrace
   \begin{array}{ll}
     \left[\left(\xi_i + a_i\right)^{\lambda_i}-1\right] /\lambda_i & \lambda_i \neq 0\\
     \ln (\xi_i + a_i) & \lambda_i = 0
   \end{array} \right.\;,
\end{equation}
with free transformation parameters $\lambda_i$ and $a_i$, where the case $\lambda_i=0$ corresponds to a log-normal transformation. For the following statements, we determine the optimal Box-Cox parameters using the procedure of maximizing the log-likelihood for $\lambda_i$ and $a_i$ explained in \cite{bib_boxcox_benjamin} and references therein.

Note that since we cannot assume the transformation parameters to be identical for each lag $i$, we need to determine the full sets $\lbrace\lambda_i\rbrace$ and $\lbrace a_i\rbrace$. There are two different ways of addressing this: Since we are, in the end, interested in the multivariate likelihood of the correlation function, the most straightforward approach is to optimize the sets of Box-Cox parameters $\lbrace\lambda_i\rbrace$ and $\lbrace a_i\rbrace$ in such a way that the full $(n+1)$-variate distribution $p(\bar\xi_0,\bar\xi_1,\ldots,\bar\xi_n)$ of the transformed variables $\bar\xi_i$  is close to a multivariate Gaussian. Alternatively, one can treat all univariate distributions $p(\xi_i)$ separately, \ie determine the optimal Box-Cox parameters in such a way that the univariate PDFs $p(\bar\xi_0),\ldots, p(\bar\xi_n)$ of the transformed variables are univariate Gaussians.

The first approach, \ie trying to Gaussianize the full $(n+1)$-variate distribution, turns out to be unsuccessful: The multivariate moments (as defined in \autoref{sec:quality_y}) of the transformed quantities $\bar{\xi}$ are hardly any different from those of the original correlation functions $\xi$. Additionally, there is barely any improvement in the Gaussianity of the univariate distributions $p(\bar{\xi_i})$ compared to $p(\xi_i)$. In contrast to that, the transformation from $\xi$ to $y$ used in our calculation of the quasi-Gaussian likelihood resulted in an improvement in skewness and kurtosis by about an order of magnitude, as described in \autoref{sec:quality_of_qg}.

The second approach, \ie treating all univariate distributions independently by trying to Gaussianize them separately, of course leads to lower univariate skewness and kurtosis in the transformed quantities $\bar{\xi}_i$; however, the multivariate moments are again almost unchanged compared to the untransformed correlation functions, indicating that this approach does not lead to a better description of the multivariate correlation function likelihood.

Thus, in summary, the quasi-Gaussian approach seems far more accurate than using Box-Cox transformations. This is not too surprising, since the latter obviously cannot properly take into account correlations between the different random variables (\ie the $\xi_i$), whereas in contrast to that, the transformation $\xi\rightarrow y$ is specifically tailored for correlation functions and thus mathematically better motivated than a general Gaussianizing method such as power transforms.

%______________________________________________________________

\section{Conclusions and outlook}
\label{sec:conclusions}
Based on the exact univariate likelihood derived in \cite{bib_david_paper} and the constraints on correlation functions derived in \cite{bib_peter_jan_paper}, we have shown how to calculate a quasi-Gaussian likelihood for correlation functions on one-dimensional Gaussian random fields, which by construction obeys the aforementioned constraints. Simulations show that the quasi-Gaussian PDF is significantly closer to the ``true'' distribution than the usual Gaussian approximation. Moreover, it is also superior to a straightforward copula approach, which couples the exact univariate PDF of correlation functions derived by \cite{bib_david_paper} with a Gaussian copula. When used in a toy-model Bayesian analysis, the quasi-Gaussian likelihood results in a noticeable change of the posterior compared to the Gaussian case, indicating its possible impact on cosmological parameter estimation.

As an outlook on future work, we would like to highlight some possible next steps: Applying the quasi-Gaussian approach to real data is obviously the ultimate goal of this project, and it would be of greatest interest to see the influence of the new likelihood on the measurement of cosmological parameters. So far, this would only be possible for 1D random fields, \eg measurements from the Ly$\alpha$ forest.
However, since most random fields relevant for cosmology are two- or three-dimensional, generalizing the quasi-Gaussian approach to higher-dimensional fields is crucial to broaden the field of applicability. As \cite{bib_peter_jan_paper} showed, the constraints on correlation functions obtained for the 1D fields are no longer optimal in higher dimensions, so the higher-dimensional constraints need to be derived first -- work in this direction is currently in progress.

%______________________________________________________________

\begin{acknowledgements}
We would like to thank Cristiano Porciani, David Keitel, and Jan Hartlap for useful help and discussion in the course of the project as well as Benjamin Joachimi for providing his Box-Cox code. We also thank our anonymous referee for constructive comments and suggestions. This work was supported by the Deutsche Forschungsgemeinschaft under the project SCHN~342/11.
\end{acknowledgements}

%______________________________________________________________

\bibliographystyle{aa}   % style aa.bst
\bibliography{bib_qgpdf_rev} % your references Yourfile.bib

%______________________________________________________________

\appendix
\section{Equivalence of the simulation methods}
\label{sec:equivalence_sims}
In this appendix, we want to prove analytically that the two simulation methods mentioned in \autoref{sec:simulations} are in fact equivalent.

The established method calculates the correlation function components $\xi_m$, $m=0,\ldots, N-1$ directly from the field components $g_n$, $n=0,\ldots, N-1$. Since we impose periodic boundary conditions on the field, this can be done using the estimator
\begin{equation}
 \xi_m=\frac{1}{N}\sum_{n=0}^{N-1} g_n\cdot g_{n+m}.
 \label{eq:proof_xi_estimator}
\end{equation}
The real-space field components are calculated from the Fourier components by
\begin{equation}
 g_i=\sum_{j=-N/2}^{N/2} \exp\left(\frac{2\pi\im}{N} ij\right)\ \tilde{g}_i,
 \label{eq:proof_g_FT}
\end{equation}
Here, we set $\tilde g_0=0$, which is equivalent to the field having zero mean in real space. Discretization and periodicity already imply $\tilde g_{N/2}\equiv \tilde g_{-N/2}$ -- still, in order not to give double weight to this mode, we set it to zero as well. Of course, we then have to do the same in our new method, \ie $P_{N/2}\equiv 0$. Note that for the sake of readability, we still include this term in the formulae of this work.

Our new method draws a realization of the power spectrum and Fourier transforms it to obtain the correlation function:
\begin{equation}
 \xi_m=\frac{2}{L}\sum_{n=1}^{N/2}P_n\cos\frac{2\pi mn}{N}.
 \label{eq:proof_xi_from_P}
\end{equation}
In both methods, the variance $\sigma^2(k_n)\equiv \sigma^2_n=\langle\left|\tilde g_n\right|^2\rangle$ of the field components in Fourier space is determined by the power spectrum, namely, for one realization,
\begin{equation}
 |\tilde{g}_k|^2=\frac{1}{L}P_k.
 \label{eq:PS_gtilde}
\end{equation}
To prove the equivalence of the two methods, we insert the Fourier transforms as given in \autoref{eq:proof_g_FT} into the estimator, \autoref{eq:proof_xi_estimator}:
\begin{eqnarray}
 \xi_m &=& \frac{1}{N}\sum_{n=0}^{N-1}\ \sum_{l=-N/2}^{N/2}\ \sum_{k=-N/2}^{N/2} \tilde{g}_l\ \exp\left(\frac{2\pi\im}{N}ln\right)\ \tilde{g}_k\ \exp\left(\frac{2\pi\im}{N}k(n+m)\right)\nonumber\\
 &=& \frac{1}{N}\sum_{n=0}^{N-1}\ \sum_{l=-N/2}^{N/2}\ \sum_{k=-N/2}^{N/2} \tilde{g}_l\ \tilde{g}_k\ \exp\left(\frac{2\pi\im}{N}(ln+kn+km)\right).
\end{eqnarray}
The sum over $n$ simply gives a Kronecker $\delta$, since
\begin{equation}
 \frac{1}{N}\sum_{n=0}^{N-1}\exp\left(\frac{2\pi\im}{N}n(l-k) \right)=\delta_{lk}.
\end{equation}
Thus, we obtain
\begin{eqnarray}
 \xi_m &=& \sum_{l=-N/2}^{N/2}\ \sum_{k=-N/2}^{N/2} \tilde{g}_l\tilde{g}_k\ \exp\left(\frac{2\pi\im}{N}km\right)\delta_{-kl}\nonumber\\
 &=& \sum_{k=-N/2}^{N/2} \tilde{g}_k\ \tilde{g}_{-k}\ \exp\left(\frac{2\pi\im}{N}km\right)\nonumber\\
 &=& \sum_{k=-N/2}^{N/2} |\tilde{g}_k|^2 \exp\left(\frac{2\pi\im}{N}km\right),
\end{eqnarray}
where in the last step, we used the fact that the $g_i$ are real, implying $\tilde g_i=\tilde g^*_{-i}$. In order to show that this is equivalent to \autoref{eq:proof_xi_from_P}, we now split the sum into two parts, omitting the zero term (since $\tilde g_0=0$, as we explained before):
\begin{eqnarray}
 \xi_m &=& \sum_{k=-N/2}^{1} |\tilde{g}_k|^2\  \exp\left(\frac{2\pi\im}{N}km\right) + \sum_{k=1}^{N/2} |\tilde{g}_k|^2 \exp\left(\frac{2\pi\im}{N}km\right)\nonumber\\
 &=& \sum_{k=1}^{N/2} |\tilde{g}_k|^2\ \exp\left(-\frac{2\pi\im}{N}km\right) + \sum_{k=1}^{N/2} |\tilde{g}_k|^2 \exp\left(\frac{2\pi\im}{N}km\right)\nonumber\\
 &=& 2\cdot \sum_{k=1}^{N/2} |\tilde{g}_k|^2 \cos\left(\frac{2\pi}{N}km\right).
\end{eqnarray}
Inserting \autoref{eq:PS_gtilde} we end up with
\begin{equation}
 \xi_m= \frac{2}{L}\cdot \sum_{k=1}^{N/2} P_k \cos\left(\frac{2\pi}{N}km\right).
\end{equation}
This is exactly the way we calculate $\xi_m$ in our new method -- thus, we proved analytically that the two methods are indeed equivalent. As mentioned before, we also confirmed this fact numerically.

\section{Analytical calculation of the \texorpdfstring{$\xi_0$}{xi0}-dependence of mean and covariance matrix}
\label{sec:analytical_xi0_dep}
As mentioned in \autoref{sec:new_approximation}, our analytical calculation of the $\xi_0$-dependence of the mean and the covariance matrix does not produce practically usable results -- nonetheless, it is interesting from a theoretical point of view and is thus presented in this appendix.

We are ultimately interested in the mean $\vec y$ and the covariance matrix $C_y$, however, we will first show calculations in $\xi$-space before addressing the problem of how to transform the results to $y$-space.

\subsection{Calculation in \texorpdfstring{$\xi$}{xi}-space}
\begin{figure*}[ht!]
      \resizebox{0.3\hsize}{!}{\includegraphics[keepaspectratio, angle=-90]{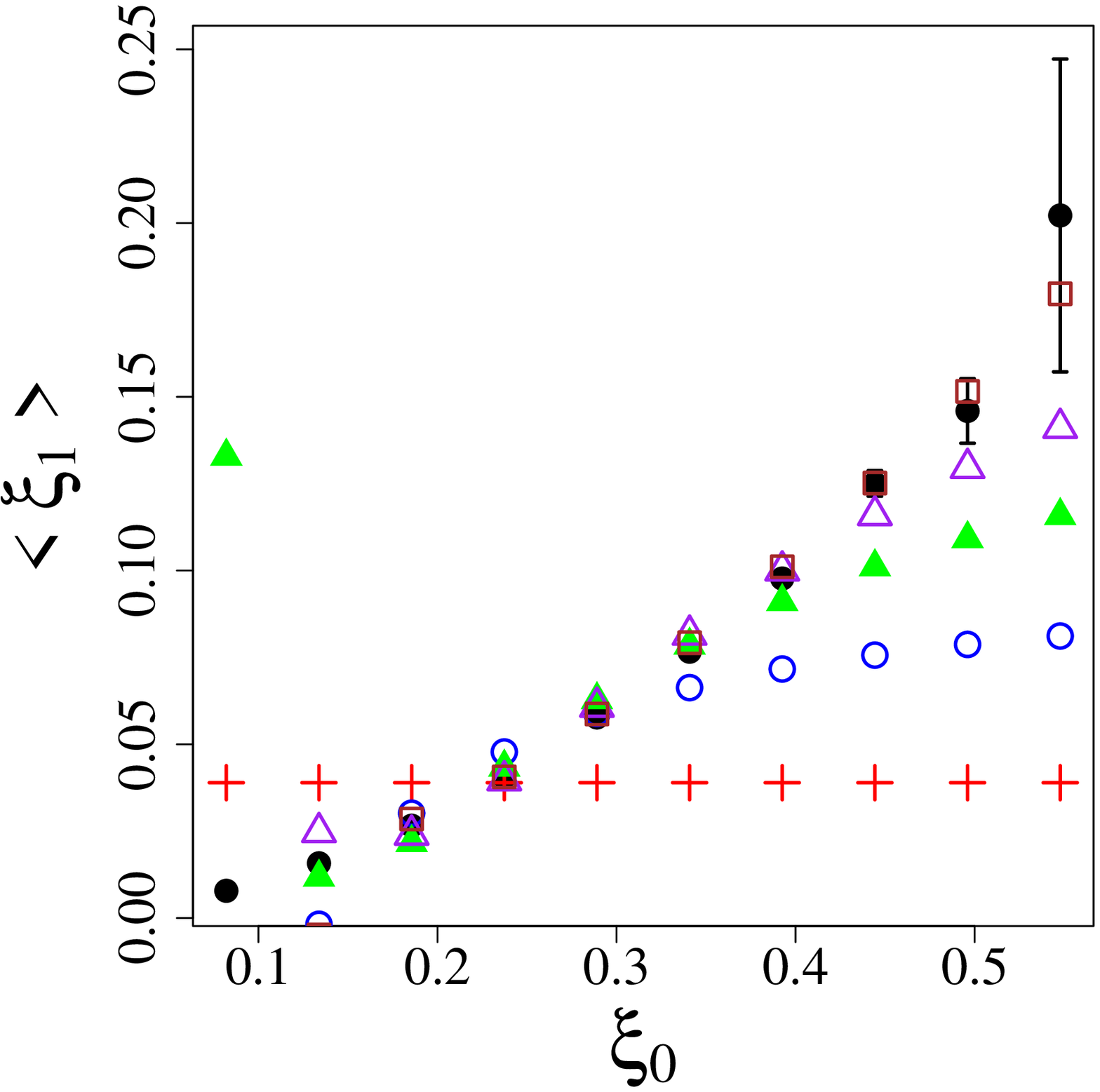}}\hfill
      \resizebox{0.3\hsize}{!}{\includegraphics[keepaspectratio, angle=-90]{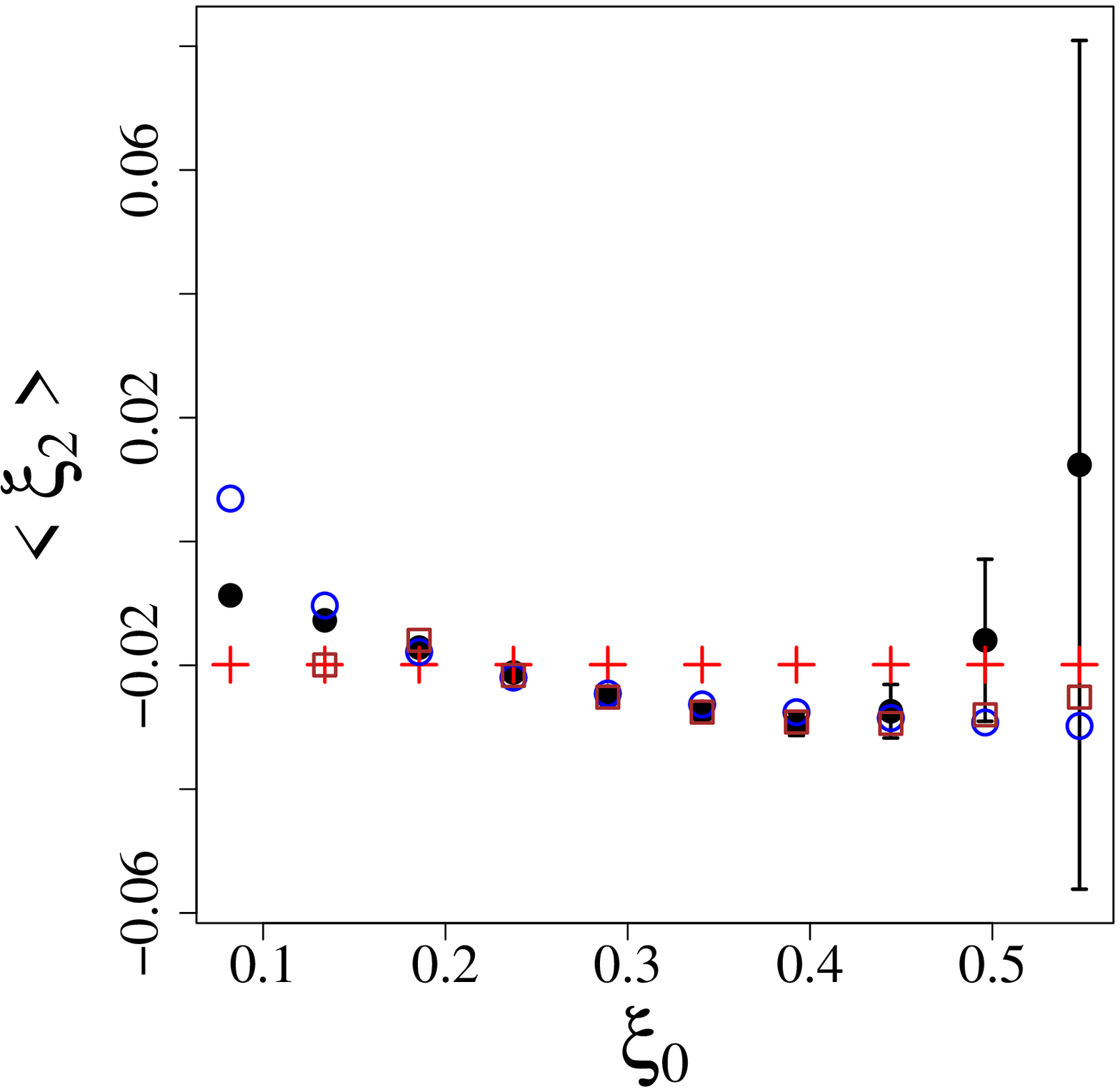}}\hfill
      \resizebox{0.3\hsize}{!}{\includegraphics[keepaspectratio, angle=-90]{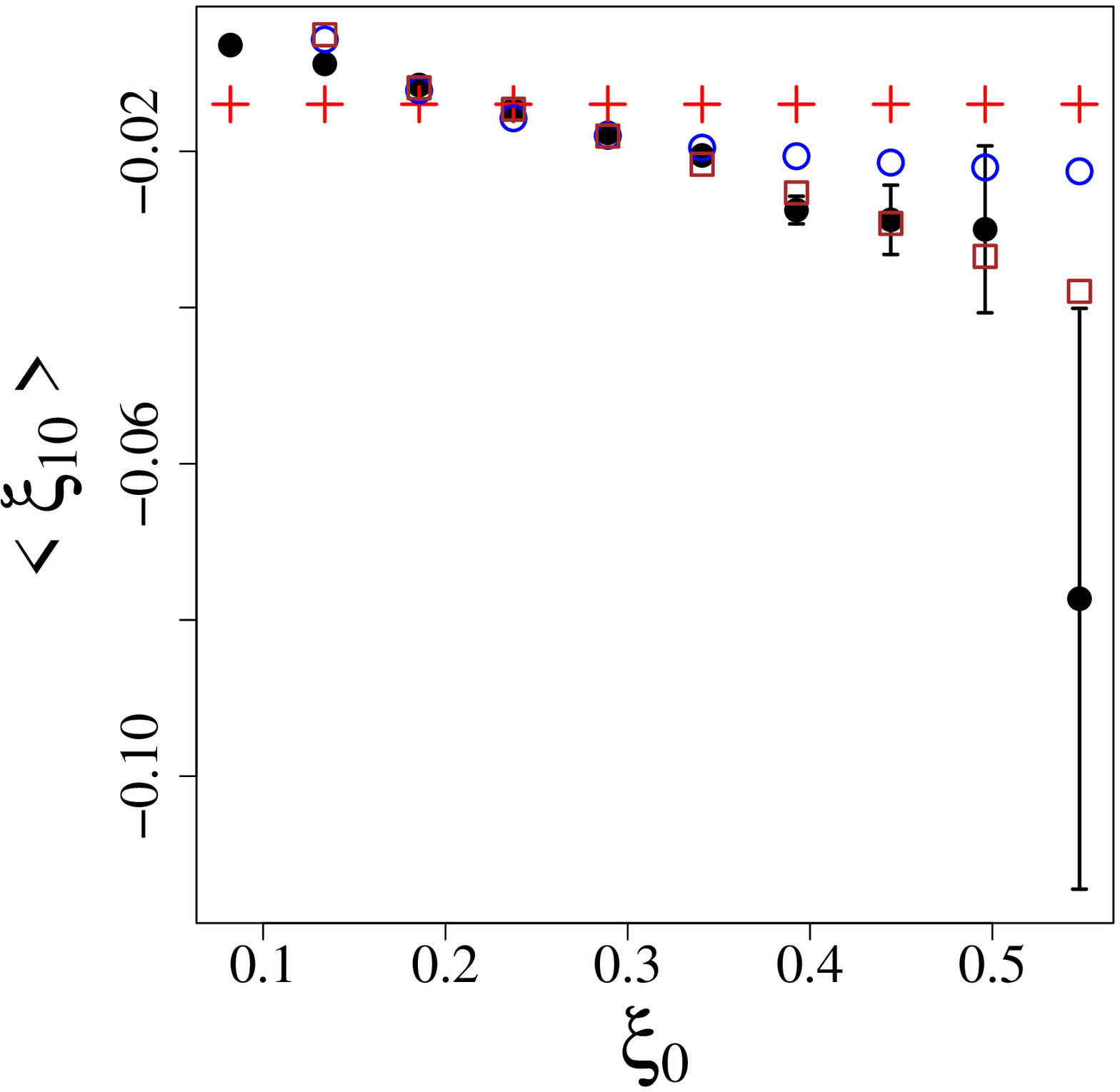}}
      \caption{The mean of $\xi_{n}$ for different $n$ as function of $\xi_0$, determined from simulations (black points with error bars) and analytically to zeroth (red crosses), first (blue circles), second (green filled triangles; left panel only), third (purple empty triangles; left panel only), and tenth (brown squares) order.}
      \label{fig:mean_ana}
\end{figure*}

The $\xi_0$-dependence of the mean $\langle\xi_1\rangle$ (where the index is purely a numbering and does not denote the lag) can be computed as
\begin{equation}
 \langle\xi_1\rangle\left(\xi_0\right) = \int\dd\xi_1\ \xi_1\ p(\xi_1|\xi_0)
 \label{eq:mean_xi_def}
\end{equation}
with the conditional probability
\begin{equation}
 p(\xi_1|\xi_0) = \frac{p(\xi_0,\xi_1)}{p(\xi_0)}.
 \label{eq:p_cond_from_joint}
\end{equation}
We define the corresponding characteristic function as Fourier transform of the probability distribution ($\Phi\leftrightarrow p$ in short-hand notation; for details on characteristic functions see \citealt{bib_david_paper},  hereafter \KS{}, and references therein):
\begin{eqnarray}
 \Phi(s_1;\xi_0) &=& \int\dd\xi_1\ \e^{\im s_1\xi_1}\ p(\xi_1|\xi_0),\\
 p(\xi_1|\xi_0) &=& \int\frac{\dd s_1}{2\pi}\ \e^{-\im s_1\xi_1}\ \Phi(s_1;\xi_0).
 \label{eq:p_cond_from_charact}
\end{eqnarray}
Making use of the characteristic function $\Psi(s_0,s_1)$ (where $\Psi(s_0,s_1)\leftrightarrow p(\xi_0,\xi_1)$) computed in \KS{}, we can also write
\begin{eqnarray}
 p(\xi_1|\xi_0) &=& \frac{1}{p(\xi_0)} \int\frac{\dd s_0}{2\pi}\int\frac{\dd s_1}{2\pi}\ \e^{-\im s_0\xi_0}\ \e^{-\im s_1\xi_1}\ \Psi(s_0,s_1)\nonumber\\
 &=& \int\frac{\dd s_1}{2\pi}\ \e^{-\im s_1\xi_1}\ \frac{1}{p(\xi_0)}\int\frac{\dd s_0}{2\pi} \e^{-\im s_0\xi_0}\ \Psi(s_0,s_1).
 \end{eqnarray}
Comparison with \autoref{eq:p_cond_from_charact} yields
\begin{equation}
 \Phi(s_1;\xi_0) = \frac{1}{p(\xi_0)} \int\frac{\dd s_0}{2\pi}\ \e^{-\im s_0\xi_0}\ \Psi(s_0,s_1).
 \label{eq:phi}
\end{equation}
Now, we can calculate the mean (\ie the first moment) from the characteristic function (equivalent to \autoref{eq:mean_xi_def} -- again, see \KS{} and references there):
\begin{eqnarray}
 \langle\xi_1\rangle(\xi_0) &=& \frac{\dd}{\im\dd s_1}\Phi(s_1;\xi_0)\ \Big|_{s_1=0}\\
 &=& \frac{1}{p(\xi_0)}\int\frac{\dd s_0}{2\pi}\ \e^{-\im s_0\xi_0}\ \frac{\dd}{\im\dd s_1}\Psi(s_0,s_1)\ \Big|_{s_1=0}.
 \label{eq:mean_xi_calc}
\end{eqnarray}
Using the result from \KS{} for the bivariate characteristic function,
\begin{equation}
 \Psi(s_0,s_1) = \prod_{n=1}^\infty\frac{1}{1-2\im s_0 \Cnzero-2\im s_1 \Cnone},
\end{equation}
where $C_{nm}=\sigma_n^2\cos(k_n x_m)$, we can calculate the derivative as
\begin{eqnarray}
 \frac{\dd}{\im\dd s_1}\Psi(s_0,s_1)\ \Big|_{s_1=0} &=& \sum_{n=1}^\infty\frac{2\Cnone}{\left(1-2\im s_0\Cnzero\right)^2}\ \prod_{k\neq n}\frac{1}{1-2\im s_0 C_{k0}}\nonumber\\
 &=& \sum_{n=1}^\infty\frac{2\Cnone}{1-2\im s_0\Cnzero}\ \prod_{k=1}^\infty\frac{1}{1-2\im s_0 C_{k0}}\nonumber\\
 &=& \Psi(s_0)\ \sum_{n=1}^\infty \underbrace{\frac{2\Cnone}{1-2\im s_0\Cnzero}}_{Y_n(s_0)},
 \label{eq:deriv_psi}
\end{eqnarray}
where we inserted the univariate characteristic function computed in \KS{},
\begin{equation}
 \Psi(s_0) = \prod_{n=1}^\infty\frac{1}{1-2\im s_0 \Cnzero}.
\end{equation}
To calculate the integral in \autoref{eq:mean_xi_calc}, we use a Taylor expansion of $Y_n(s_0)$ from \autoref{eq:deriv_psi}:
\begin{equation}
 Y_n(s_0) \approx \sum_{k=0}^\infty 2^{k+1}\ \left(\im s_0\right)^k\ \Cnzero^k\Cnone.
 \label{eq:taylorYn}
\end{equation}
We insert the derivative into \autoref{eq:mean_xi_calc} and thus obtain
\begin{eqnarray}
 \langle\xi_1\rangle\left(\xi_0\right) \approx \frac{1}{p\left(\xi_0\right)}\int\frac{\dd s_0}{2\pi}\ \e^{-\im s_0\xi_0}\ \Psi\left(s_0\right)\nonumber\\
 \sum_{n=1}^\infty\ \sum_{k=0}^\infty 2^{k+1}\ \left(\im s_0\right)^k\ \Cnzero^k\Cnone.
 \label{eq:mean_xi_calc2}
\end{eqnarray}
According to the definition of $\Psi(s_0) \leftrightarrow p(\xi_0)$,
\begin{equation}
 \frac{\dd^k p\left(\xi_0\right)}{\dd\xi_0^k} = \int\frac{\dd s_0}{2\pi}\left(-\im s_0\right)^k\ \e^{-\im s_0\xi_0}\ \Psi\left(s_0\right),
\end{equation}
and thus, after changing the order of summation and integration, \autoref{eq:mean_xi_calc2} can finally be written as
\begin{equation}
 \langle\xi_1\rangle(\xi_0) = \sum_{k=0}^\mathrm{order}\ \sum_{n=1}^\mathrm{modes} 2^{k+1}\ \Cnzero^k\Cnone\ (-1)^k\ \frac{\dd^k p(\xi_0)}{\dd\xi_0^k}\ \frac{1}{p(\xi_0)}.
 \label{eq:mean_xi_final}
\end{equation}
Inserting the known result for $p(\xi_0)$ and calculating its derivatives allows us to compare the analytical result to simulations. The results can be seen in \autoref{fig:mean_ana}; here, the black points with error bars show the mean of $\xi_n$ for different lags $n$ as determined from simulations ($100\ 000$ realizations, Gaussian power spectrum with $L k_0=100$), and the colored symbols show the analytical results to different order (see figure caption). It seems that, although the Taylor series in \autoref{eq:mean_xi_final} does not converge, a truncation at order 10 yields sufficient accuracy, barring some numerical issues for very low $\xi_0$-values.

\begin{figure*}[ht!]
      \resizebox{0.3\hsize}{!}{\includegraphics[keepaspectratio, angle=-90]{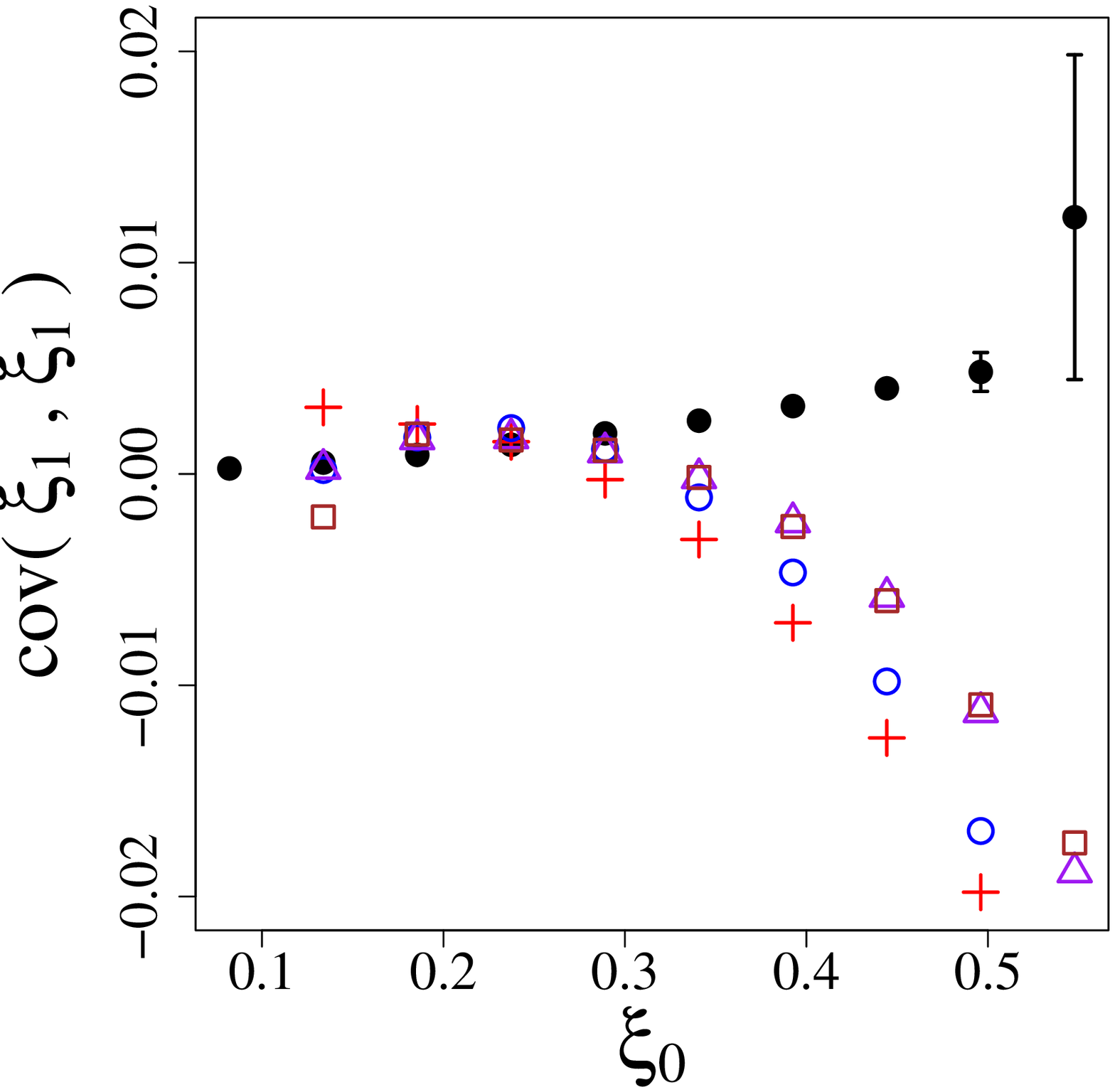}}\hfill
      \resizebox{0.3\hsize}{!}{\includegraphics[keepaspectratio, angle=-90]{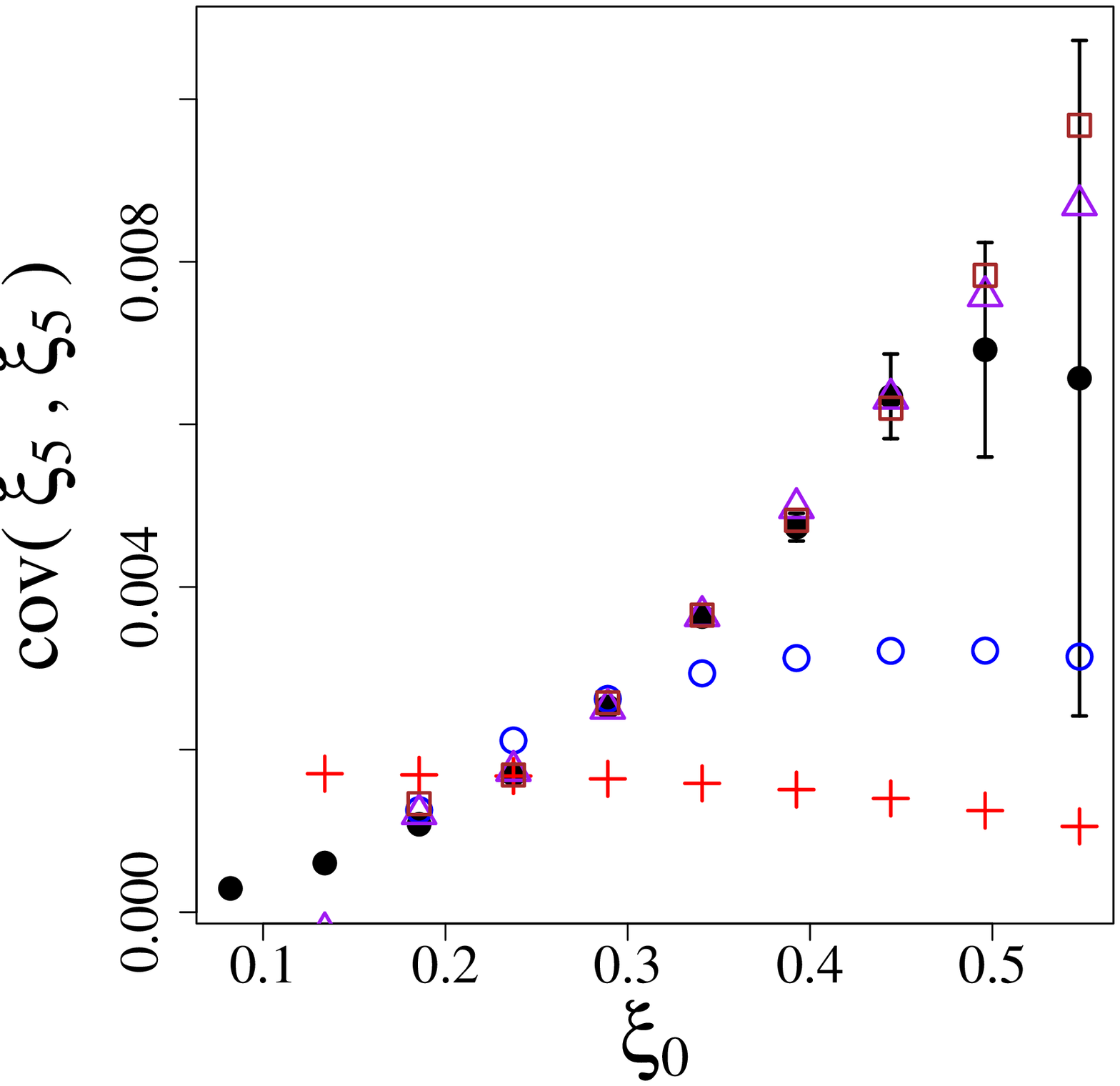}}\hfill
      \resizebox{0.3\hsize}{!}{\includegraphics[keepaspectratio, angle=-90]{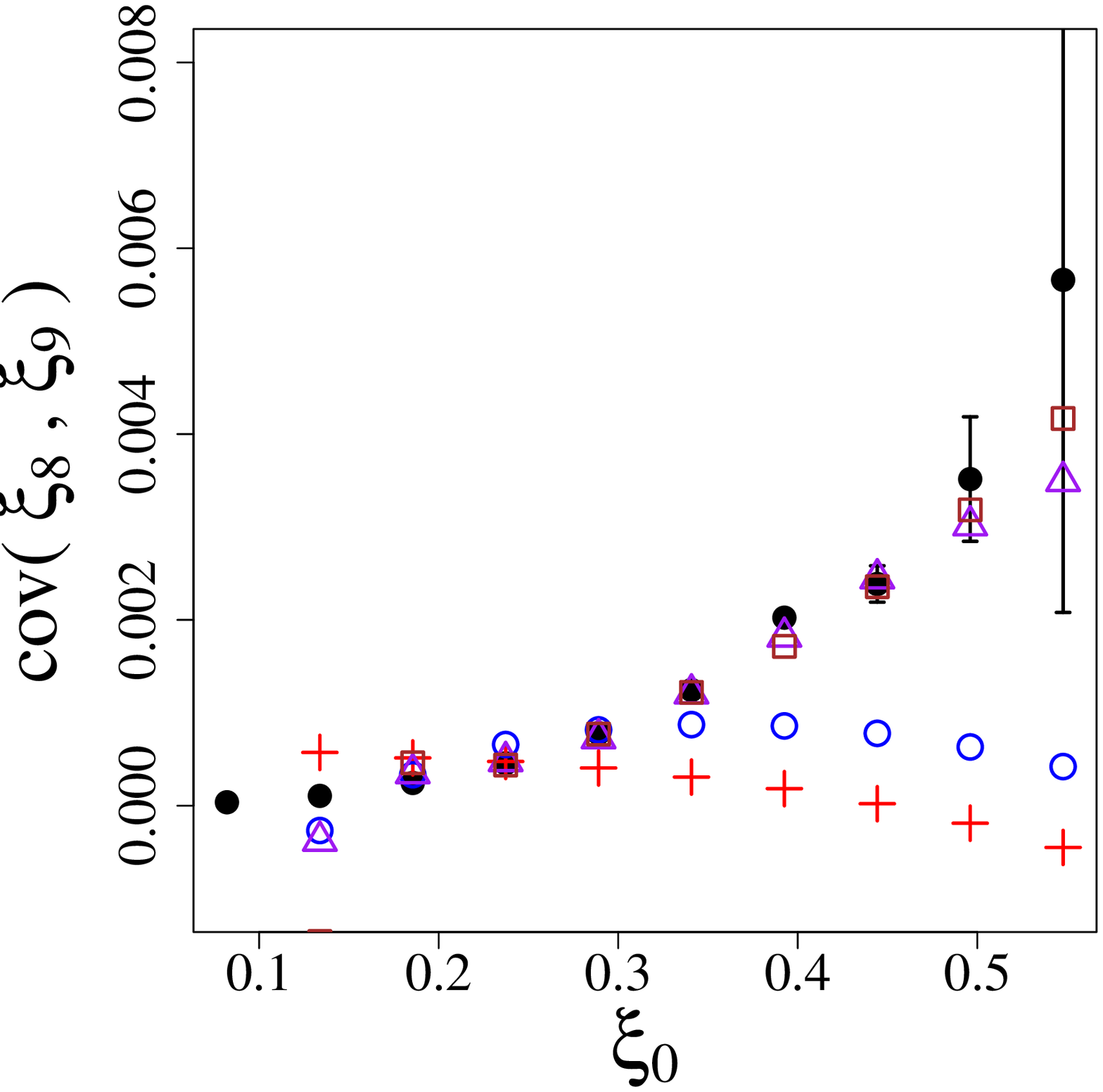}}
      \caption{Different elements of the covariance matrix $C(\lbrace \xi_n\rbrace)$, determined from simulations (black points with error bars) and analytically to zeroth (red crosses), first (blue circles), fifth (purple triangles), and tenth (brown squares) order.}
      \label{fig:cov_ana}
\end{figure*}

The $\xi_0$-dependence of the covariance matrix $C_\xi$ can be calculated in a similar way. We start from the general definition of covariance,
\begin{eqnarray*}
 \cov\left(\xi_1,\xi_2\right)\left(\xi_0\right) &\equiv& \langle\left(\xi_1-\langle\xi_1\rangle\right)\ \left(\xi_2-\langle\xi_2\rangle\right)\rangle_{\xi_0}\nonumber\\
 &=& \int\dd\xi_1\ \dd\xi_2\ (\xi_1-\langle\xi_1\rangle)\ (\xi_2-\langle\xi_2\rangle)\ p(\xi_1,\xi_2|\xi_0)\nonumber\\
 &=& \underbrace{\int\dd\xi_1\ \dd\xi_2\ \xi_1\xi_2\ p(\xi_1,\xi_2|\xi_0)}_{\equiv A}\nonumber\\
 &+& \langle\xi_1\rangle\left(\xi_0\right)\ \langle\xi_2\rangle\left(\xi_0\right) \underbrace{\int\dd\xi_1\ \dd\xi_2\ p(\xi_1,\xi_2|\xi_0)}_{=1}\nonumber\\
 &-& \langle\xi_1\rangle\left(\xi_0\right) \underbrace{\int\dd\xi_1\ \dd\xi_2\ \xi_2\ p(\xi_1,\xi_2|\xi_0)}_{=\langle\xi_2\rangle\left(\xi_0\right)} \nonumber\\
 &-& \langle\xi_2\rangle\left(\xi_0\right) \underbrace{\int\dd\xi_1\ \dd\xi_2\ \xi_1\ p(\xi_1,\xi_2|\xi_0)}_{=\langle\xi_1\rangle\left(\xi_0\right)}.
\end{eqnarray*}
The integral $A$ can again be expressed in terms of the characteristic function $\Phi(s_1,s_2;\xi_0)\leftrightarrow p(\xi_1,\xi_2|\xi_0)$:
\begin{equation}
 A = \frac{\dd}{\im\dd s_1}\frac{\dd}{\im\dd s_2}\ \Phi(s_1,s_2;\xi_0) \Big|_{s_1=s_2=0}
\end{equation}
Similar to the previous calculations,
\begin{equation}
 \Phi\left(s_1,s_2;\xi_0\right) = \frac{1}{p\left(\xi_0\right)} \int\frac{\dd s_0}{2\pi}\ \e^{-\im s_0\xi_0}\ \Psi\left(s_0,s_1,s_2\right)
\end{equation}
with the trivariate characteristic function
\begin{equation}
 \Psi\left(s_0,s_1,s_2\right) = \prod_{n=1}^\infty\frac{1}{1-2\im s_0 \Cnzero-2\im s_1 \Cnone-2\im s_2 \Cntwo}.
 \label{eq:Psi3}
\end{equation}
Calculating the second derivative of \ref{eq:Psi3} yields
\begin{eqnarray}
 \frac{\dd}{\im\dd s_1}\frac{\dd}{\im\dd s_2} &\Psi& \left(s_0,s_1,s_2\right)\Big|_{s_1=s_2=0} \nonumber\\
 &=& \Psi\left(s_0\right) \left[\sum_{n=1}^\infty \underbrace{\frac{4\Cnone\Cntwo}{(1-2\im s_0\Cnzero)^2}}_{Z_n(s_0)} \right.\nonumber\\
 &+& \left. \sum_{n,k=1}^\infty \underbrace{\frac{4\Cnone\Cktwo}{(1-2\im s_0\Cnzero)(1-2\im s_0\Ckzero)}}_{Z_{k,n}(s_0)} \right].
\end{eqnarray}
The Taylor expansions of $Z_n(s_0)$ and $Z_{k,n}(s_0)$ read
\begin{eqnarray}
 Z_n(s_0) &\approx& \sum_{k=0}^\infty 2^{k+2}\ (k+1)\left(\im s_0\right)^k\ \Cnzero^k\Cnone\Cntwo,\\
 Z_{k,n}(s_0) &\approx& \sum_{l=0}^\infty 2^{l+2}\ \left(\im s_0\right)^l\ \Cktwo \Cnone \sum_{p=0}^{l}\Ckzero^p\ \Cnzero^{l-p}.
\end{eqnarray}
Using it as well as the expansion \ref{eq:taylorYn}, we finally obtain
\begin{eqnarray}
 \cov\left(\xi_1,\xi_2\right)\left(\xi_0\right) &=& \sum_{k=0}^\mathrm{order}\ \frac{1}{p(\xi_0)}\frac{\dd^k p\left(\xi_0\right)}{\dd\xi_0^k} \left[ \sum_{n}^\mathrm{modes} (-1)^k\ 2^{k+2}\ (k+1)\right.\nonumber\\
 &\times& \Cnzero^k\Cnone\Cntwo + \sum_{n,m}^\mathrm{modes} (-1)^k\ 2^{k+2}\ \Cmtwo\ \Cnone\nonumber\\
 &\times& \left. \sum_{p=0}^{k}\Cmzero^p\ \Cnzero^{k-p} \right] - \langle\xi_1\rangle(\xi_0)\ \langle\xi_2\rangle(\xi_0).
  \label{eq:cov_xi_final}
\end{eqnarray}
We show a comparison of the results (for different elements of the covariance matrix) from simulations and the analytical formula in \autoref{fig:cov_ana}. Again, the black dots are obtained from simulations and the colored symbols represent the results from \autoref{eq:cov_xi_final}, where the last term (\ie the one containing the mean values $\langle\xi_n\rangle$) was calculated up to tenth order, thus providing sufficient accuracy, as previously shown. As before, there are some numerical problems for very small values of $\xi_0$. Additionally, the analytical results do not agree with the simulations for small lags, as can be seen from the left-most panel (the same holds for other covariance matrix elements involving small lags). However, for the the higher-lag examples (\ie the right two panels), a truncation of the Taylor series at tenth order seems to be accurate enough.

\subsection{Transformation of mean and covariance matrix to \texorpdfstring{$y$}{y}-space}
In the previous section, we showed how to calculate the ($\xi_0$-dependent) mean and covariance matrix in $\xi$-space. The computation of the quasi-Gaussian approximation, however, requires the mean $\vec y$ and the covariance matrix $C_y$ in $y$-space, which cannot be obtained from those in $\xi$-space in a trivial way due to the highly non-linear nature of the transformation $\xi\rightarrow y$.

Thus, instead of settling for a linear approximation, we have to choose a more computationally expensive approach. Namely, we calculate the first and second moments (in $\xi$) of the quasi-Gaussian distribution as functions of the mean and (inverse) covariance matrix in $y$-space and equate the result to the analytical results, \ie we solve a set of equation of the form
\begin{eqnarray}
 \int \dd\xi\ \xi_i\ p\left(\vec\xi; \langle\vec y\rangle, C^{-1}_{y}\right) &=& \langle\xi_i\rangle_\mathrm{ana}\\
 \int \dd\xi\ \xi_j\ \xi_k\ p\left(\vec\xi; \langle\vec y\rangle, C^{-1}_{y}\right) &=& \langle\xi_j\ \xi_k\rangle_\mathrm{ana},
\end{eqnarray}
where we did not write down the $\xi_0$-dependence explicitly for the sake of readability. 

Note that this is a complicated procedure, since the integration on the equations' left-hand sides can only be performed numerically (we make use of a Monte-Carlo code from \citealt{bib_nr}). In order to solve the equation set (consisting of $N+\frac{1}{2}N(N+1)$ equation for an $N$-variate distribution) we use a multi-dimensional root-finding algorithm (as provided within the GSL, \citealt{bib_gsl}). However, due to the high dimensionality of the problem, this procedure does not seem practical, since it is computationally very expensive -- in addition to that, any possible gain in accuracy is averted by the required heavy use of purely numerical methods.

Thus, as described in \autoref{sec:new_approximation}, we refrain from using our analytical results for the mean and covariance matrix and simply determine them (as well as their $\xi_0$-dependence) from simulations, which we have shown to be sufficiently accurate.

\end{document}